\documentclass[journal]{IEEEtran}
\usepackage{upgreek}
\usepackage{graphicx}
\usepackage{subfigure}
\usepackage{amsmath,bm}
\usepackage{amsfonts,amssymb}
\usepackage{enumerate}
\usepackage[mathscr]{euscript}
\usepackage[square, comma, sort&compress, numbers]{natbib}
\usepackage{color}
\usepackage{url}
\usepackage{soul}

\usepackage{mathtools, cuted}
\usepackage[english]{babel}
\usepackage{lipsum}
\usepackage{multirow}
\usepackage{diagbox}
\usepackage{makecell}
\usepackage[flushleft]{threeparttable}

\ifCLASSINFOpdf
\else
\fi

\begin{document}

\title{Binocular Localization Using Resonant Beam}
\author{\normalsize Mengyuan Xu,  Mingqing Liu, Qingwei Jiang, Wen Fang, Qingwen Liu,~\IEEEmembership{Senior Member,~IEEE} and

Shengli Zhou~\IEEEmembership{Fellow,~IEEE}


\thanks{
	M. Xu, M. Liu, Q. Jiang and Q. Liu
	are with the College of Electronics and Information Engineering, Tongji University, Shanghai 201804, China
	(e-mail: xumy@tongji.edu.cn, clare@tongji.edu.cn, jiangqw@tongji.edu.cn, qliu@tongji.edu.cn).}
\thanks{
	W. Fang is with School of Electronic Information and Electrical Engineering, Shanghai Jiao Tong University, Shanghai 200240, China (e-mail: wendyfang@sjtu.edu.cn).
}
\thanks{
    S. Zhou is with Department of Electrical and Computer Engineering, University of Connecticut, Storrs, CT 06250, USA (e-mail: shengli.zhou@uconn.edu).
}
	
}

\maketitle

\begin{abstract}
Locating mobile devices precisely in indoor scenarios is a challenging task because of the signal diffraction and reflection in complicated environments. One vital cause deteriorating the localization performance is the inevitable power dissipation along the propagation path of localization signals. In this paper, we propose a high-accuracy localization scheme based on the resonant beam system (RBS) and the binocular vision, i.e., binocular based resonant beam localization (BRBL). The BRBL system utilizes the energy-concentrated and self-aligned transmission of RBS to realize high-efficiency signal propagation and self-positioning for the target. The binocular method is combined with RBS to obtain the three-dimensional (3-D) coordinates of the target for the first time. To exhibit the localization mechanism, we first elaborate on the binocular localization model, including the resonant beam transmission analysis and the geometric derivation of the binocular method with RBS. Then, we establish the power model of RBS, and the signal and noise models of beam spot imaging, respectively, to analyse the performance of the BRBL system. Finally, the numerical results show an outstanding performance of centimeter level accuracy (i.e., $<5\mathrm{cm}$ in $0.4\mathrm{m}$ width and $0.4\mathrm{m}$ length effective range at $1\mathrm{m}$ vertical distance, $<13\mathrm{cm}$ in $0.6\mathrm{m}$ width and $0.6\mathrm{m}$ length effective range at $2\mathrm{m}$ vertical distance), which applies to indoor scenarios.

\end{abstract}



\section{Introduction}
\label{sec:intro}

\IEEEPARstart{W}{ith} the emergence of diversified mobile intelligence devices, e.g., augmented reality (AR), virtual reality (VR), etc., and the enhancement of 5th generation (5G) mobile network, location based services (LBS) have attracted scientific and corporate interest for its huge practical value and market opportunity~\cite{dey2001understanding}. In outdoor scenarios, there are already mature solutions such as the Global Positioning System (GPS)~\cite{GPS}. Nevertheless, indoor localization system (ILS) requires higher accuracy~\cite{8732442} and is more challenging~\cite{alarifi2016ultra}, which has promoted extensive research around the world. Plenty of solutions to accurate indoor localization have been proposed~\cite{jekabsons2011analysis}. 
	
	Two typical classifications of these solutions are formed according to the types of wireless technology and the types of localization schemes. Several types of wireless technologies are used for indoor location, such as Radio Frequency (RF) Signals based localization, Light-based localization, Sound-based localization, etc.~\cite{brena2017evolution}. As for the localization scheme, indoor location sensing generally contains triangulation algorithm, scene analysis and proximity~\cite{liu2007survey}. Triangulation is the most widely studied scheme, and is subdivided by specific lateration techniques, e.g. received signal strengths (RSS), time of arrival (TOA) or time difference of arrival (TDOA), and angulation techniques, i.e., angles of arrival (AOA).


%
\begin{figure}[!t]
    \centering
     \includegraphics[scale=0.45]{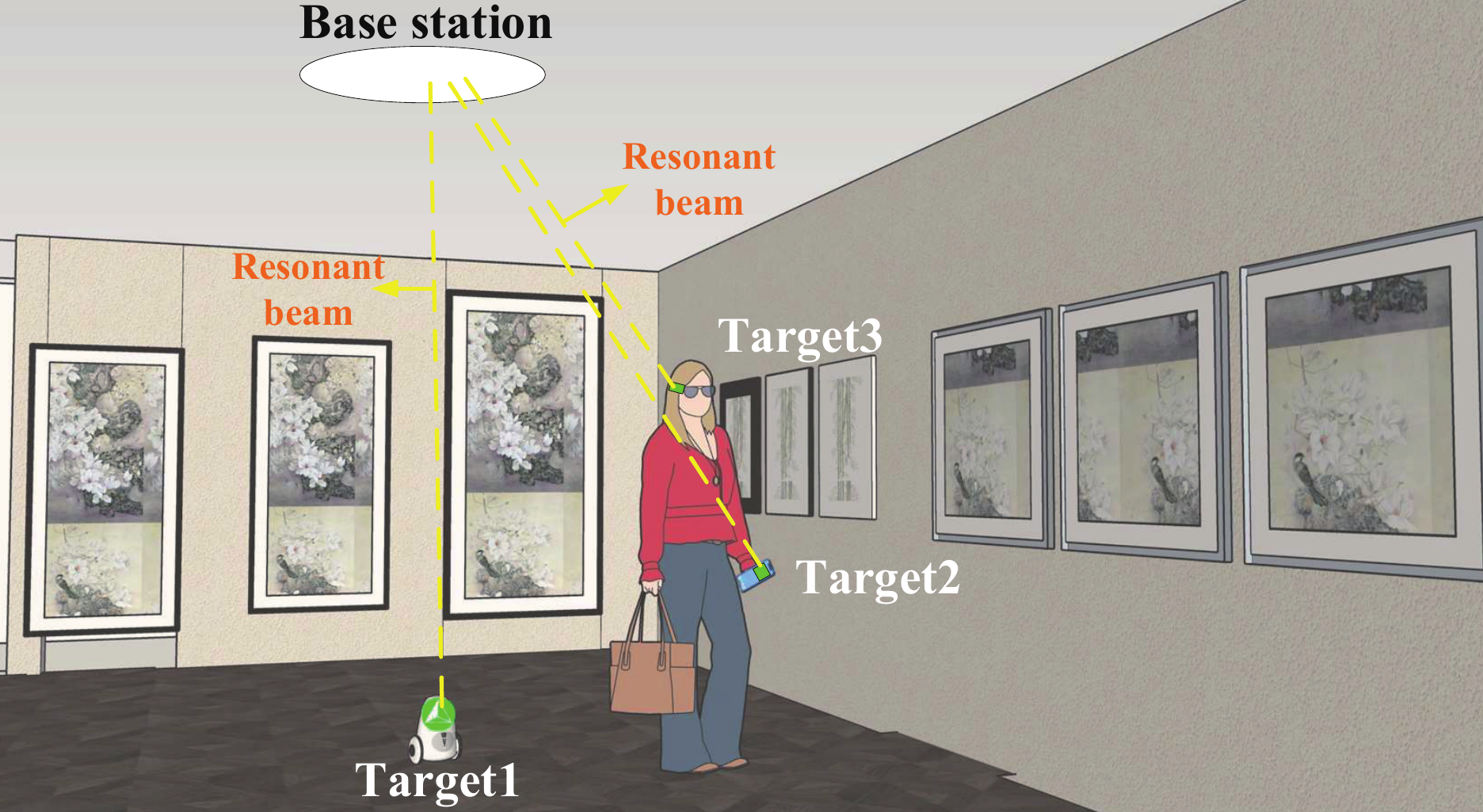}
    \caption{A typical application scenario of the BRBL system.}
    \label{Fig.Scenario}
\end{figure}

RF-based localization systems are widely investigated, including Wi-Fi, ZigBee, Bluetooth low energy (BLE), RFID and ultra-wideband (UWB). One attractive benefit of RF-based ILSs is the lower expenses. Because of the mature infrastructure, the installation cost for most RF-based technologies (except for UWB) is low. While the safety issues, shortage of spectrum resource and environmental disturbance~\cite{app11010279} stand in the way of more real progress. 

Another popular research topic is light-based localization. Infrared (IR) light is one early option, for instance, a pioneering localization system ``Active Badge''~\cite{want1992active}, which locates the IR tag by proximity in an IR sensor network. Visible light, especially LED lights, is a promising technology for indoor localization~\cite{hassan2015indoor}. Hence, a lot of works that exploit RSS~\cite{zhou2012indoor,won2013three,ComparisonRSS}, TDOA~\cite{nadeem2014highly,panta2012indoor}, or AOA~\cite{ComparisonAOA,prince2012two} algorithms in visible light ILSs are available. However, low directivity, power divergence, and inconsistent modulation schemes are still bottlenecks of the light-based localization technologies~\cite{brena2017evolution}. It is still necessary to create more accurate channel models (e.g., diffuse propagation models)~\cite{hassan2015indoor} if the localization scheme relies on the received power.


Resonant beam system (RBS) is an innovative technology for over-the-air power transfer as well as wireless communication, which was first put forward in~\cite{Qliu}. The primary structure of RBS is an active resonator formed by two spatially separated retroreflectors and a crystal material acting as the gain medium. Specifically, one retroreflector and the crystal constitute a transmitter (TX), while the other retroreflector acts as a receiver (RX). Resonant beam (i.e., intra-cavity laser beam) traveling within the resonator carries the power as well as signals. RBS is also known as resonant beam charging (RBC) in field of wireless power transfer (WPT), which is demonstrated capable of long-range, high-power and safe WPT~\cite{wangwei2019,fang2021endtoend,fang2021safe}. Besides, the high-performance communication designs proposed in~\cite{xiongRBComTWC,xiongRBComIOTJ} based on RBS can be uniformly termed as resonant beam communication (RBCom)~\cite{xiongRBComMag}, which is regarded as a promising communication solution~\cite{lmq2021}.

Concerning the technical advantages of RBS, we here propose a 3-dimensional (3-D) localization method utilizing resonant beam, i.e., binocular based resonant beam localization (BRBL). The BRBL is characterized with two identical TXs integrated in the base station (BS) and an RX integrated in the battery-free target. Owing to the collimation property of laser, the incident angles of the resonant beam reflect the orientation of the target. Afterwards, by processing the received beam at two TXs and applying the binocular localization algorithm, we can estimate the target's 3-D location.

Resonant beam is spontaneously established once the threshold condition is met. Thus in BRBL, two wireless links between the fixed BS and mobile target are automatic built, free of active emission from the target, if the target is within the field of view (FoV) of the BS. Besides, due to the structure of repeated feedback and gain, the power gets preserved within the cavity and undergoes amplification multiple times. In other words, the transmission links in BRBL are highly energy-concentrated, so that high accuracy location estimation can be achieved. The scheme of BRBL is a combination of vision analysis and AOA triangulation calculation. Coupled with the binocular algorithm, BRBL has advantages for its no need for multiplexing and easier implementation. So BRBL is a promising design and Fig.~\ref{Fig.Scenario} shows an example application scenario in an indoor exhibition hall. In this scenario, the BS is installed on the ceiling and provides location services for the targets within its effective coverage, e.g., the moving robot on the floor, the smartphone and the intelligence glasses the visitor wearing, etc.

The contributions of this paper are:

\begin{itemize}
    \item [c1)] We propose a BRBL system utilizing the energy-concentrated and self-aligned transmission of RBS for high-accuracy localization without active emission of the target. Besides, to the best of our knowledge, it is the first proposal to apply the binocular method in RBS for position estimation, so that the 3-D coordinate can be obtained directly without other embedded systems.
    \item [c2)] We establish an analytical model of the proposed BRBL system relying on the combination of binocular localization method and the RBS energy transfer model, based on which the numerical results and comparative analysis demonstrates that the proposed localization system outperforms the existing schemes with binocular vision mechanism and high precision, e.g., centimeter level accuracy.
\end{itemize}

This paper’s structure is as follows: after this introduction, we briefly present the structure design of BRBL as well as the operating principle in Section \uppercase\expandafter{\romannumeral2}. In Section \uppercase\expandafter{\romannumeral3}, we build mathematical models of the localization system from monocular imaging to binocular localization. In Section \uppercase\expandafter{\romannumeral4}, we analyse the system errors based on the beam power cycle in the resonator and the linear signal model with various noises. With the analytical model built, the simulation results are exhibited in Section \uppercase\expandafter{\romannumeral5}. Finally, we conclude in Section \uppercase\expandafter{\romannumeral6}.

\label{sec:overview}
\begin{figure}[!t]
    \centering
     \includegraphics[scale=0.88]{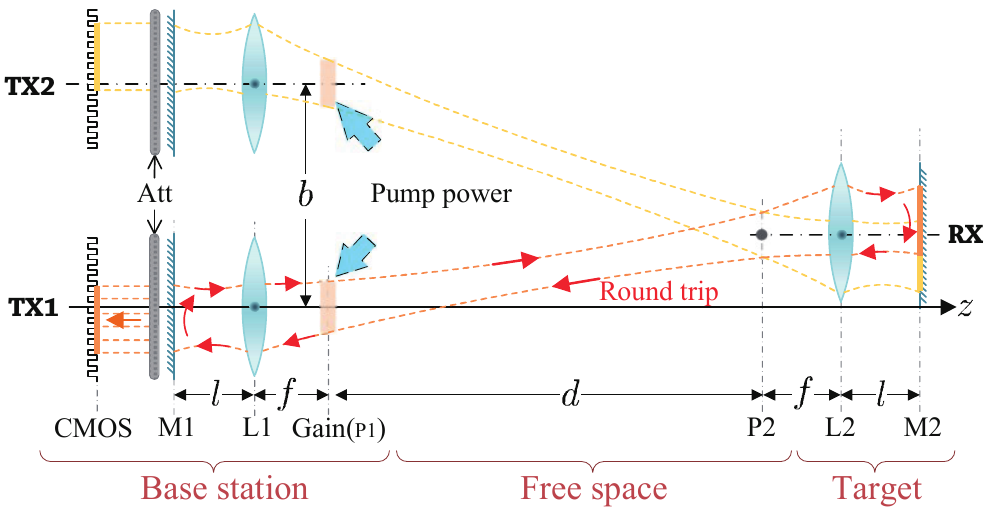}
    \caption{The structure and localization mechanism of BRBL.}
    \label{Fig.RBS}
\end{figure}
\section{System Architecture}

Based on the self-aligned resonant beam, BRBL is characterized by the high-accuracy location estimation without beam steering control. In this section, we depict the structure design of the proposed BRBL system. Moreover, we elaborate on the binocular method as well as the resonant beam transmission process to reveal the primary principle of the localization.

\subsection{Structure Design}
As depicted in Fig.~\ref{Fig.RBS}, in BRBL, we adopt two identical TXs in the BS and a built-in RX in the target. Each TX consists of a gain medium, a retroreflector, an attenuation piece (i.e., Att), and an image sensor (i.e., Complementary Metal Oxide Semiconductor, CMOS). The RX is constituted only by a retroreflector. The distance between these two TXs is depicted as $b$. A TX and an RX jointly form a resonator as a basic RBS structure, where the resonant beam transfers over the air between the separated transceivers. Thus, if the target is located within the effective scope of the BS, each of the two TXs in BS can separately form a resonator with the RX in target. Resonant beam emerges and oscillates back and forth in each resonator. The principles and features have been well-documented in~\cite{qqzhang,fang2021endtoend}, and we detail two specific structure designs as follows:
\begin{itemize}
    \item [d1)] The retroreflector that we adopt in both TX and RX is a cat's eye retroreflector, consisting of a mirror and a focal lens, e.g., M1 and L1 in the TX (or M2 and L2 in the RX). Besides, the interval $l$ between the mirror and the lens is a little larger than the focal length $f$ of the focal lens, ensuring the stability of the resonator~\cite{xiongRBComTWC} so that the resonant beam will not escape from the cavity.
    \item [d2)] The gain medium is fixed at the pupil of the cat's eye retroreflector in each TX so that the resonant beam can pass through it. The travel of the resonant beam in the cavity is a repetitious round trip, passing through the pupil twice in a loop. Thus, the beam experiences adequate amplification with the gain medium fixed at the pupil when oscillating within the cavity.
\end{itemize}

 Moreover, the mirror at the end of TX is partially transparent to leak out a small fraction of resonant beam impinging on the CMOS. The CMOS then converts the beam to digital values and exports beam spot images for the target position estimation. Considering the limited full well capacity of CMOS, we also paste an Att behind the mirror in TX to reduce the output resonant beam intensity.


\subsection{Operating Principle}
In each resonator, two opposing retroreflectors and a gain medium jointly perform the function of highly selective feedback. The resonant beam is stimulated from the gain medium and kept within the cavity once the laser oscillation condition is satisfied. As in Fig.~\ref{Fig.RBS}, the power pumped on the gain medium provides the energy source, triggers the population inversion and stimulates the photons emission. Photons not escaping from the resonator then transfer through each element within the resonator and experience transmission losses and output coupling losses. The stable oscillation is built once the amplification in the gain medium compensates for the total loss of the beam transmission in the resonator. Hence, in the proposed structure, the mirrors in retroreflectors are highly-reflective.

During the transmission of the resonant beam in a loop, it goes through different optical elements as well as free space. In our system, the resonant beam transmission process can be expressed briefly as: beam propagates from the gain medium to the retroreflector in RX along the $+z$-axis direction, then is reflected back to the medium along the $-z$-axis direction, experiences amplification in the medium, then propagates towards the retroreflector in TX, finally gets reflected back to the gain medium in the same direction as the primary direction. For the released portion of the beam from the mirror at the BS, it goes through individual diffraction over the air and finally strikes the CMOS.


Different areas of the CMOS will be illuminated in each TX, which is relative to the orientation of the incident beam. From single image exported from one TX, we can estimate two incident angles, i.e., elevation angle and azimuth angle, according to the distribution of beam spots. Besides, in BRBL, the BS is often installed on the ceiling of a room as in Fig.~\ref{Fig.Scenario}, thus the elevation angle here is the angle between the incident beam and the aligned vertical line. Then, four angle parameters can be estimated from the two images, based on which we can derive the precise 3-D localization of the target.


\section{Binocular Localization Model}
\label{sec:datamodel}
In this section, the binocular localization model is established, including the monocular imaging principle for single pair of transceiver, the 3-D coordinates algorithm using binocular localization, and the calculation of effective localization scope.

\subsection{Monocular Imaging}
In BRBL, the beam spot positions on the two CMOSs embedded in the BS will change according to the target's position. Thus, to derive the target's position with the binocular localization method, we should obtain the beam field distribution on each CMOS with the target at an arbitrary position and calculate the beam spot position change first. 

According to Huygen's principle and diffraction integral, the transmission process can also be theoretically expressed as: an initial field distribution starts from a given $x$-$y$-plane, goes through diffraction process described above, and keeps iterating until self-reproducing mode appears~\cite{hodgson2005laser}. Additionally, we apply the self-reproducing mode theory to explain the formation of stable resonant beam.

\begin{figure}[!t]
	\centering
	\includegraphics[scale=1]{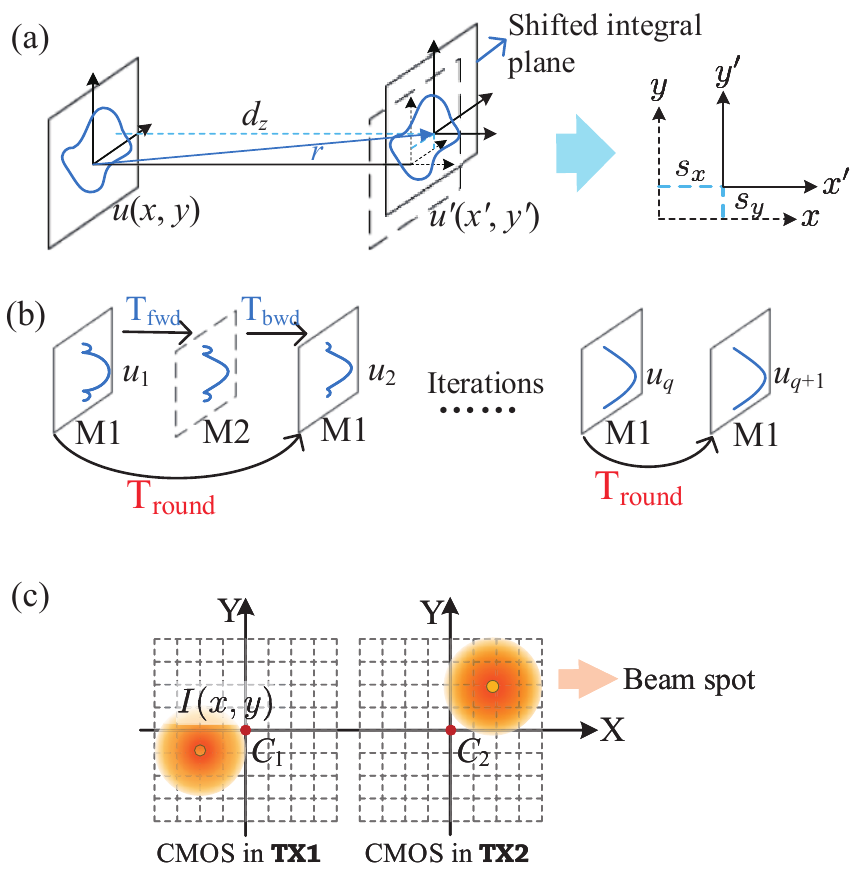}
	\caption{(a) Beam diffraction calculation between two non-coaxial planes, (b) Self-reproducing process of the beam field and (c) Beam spots impinging on the CMOSs}
	\label{Fig.iteration}
\end{figure}

The transmission of an initial beam field distribution $u(x,y)$ from one plane to the other plane can be depicted by operation $\mathrm{T}(u)$~\cite{FFT} as:
\begin{equation}
\begin{aligned}
   &\mathrm{T}(u) =\mathcal{F}^{-1}\left\{\mathcal{F}\{u(x,y)\} H\left(\nu_{x}, \nu_{y}, d_z\right)\right\}, \\
   & H\left(\nu_{x}, \nu_{y}, d_z\right)=\exp \left(i 2 \pi d_z \sqrt{\frac{1}{\lambda^2}-\nu_{x}^{2}-\nu_{y}^{2}}\right), \\
\end{aligned}
\end{equation}
where $\mathcal{F}$ and $\mathcal{F}^{-1}$ denote fast Fourier transfer and inverse Fourier transfer, respectively; $H\left(\nu_{x}, \nu_{y}, d_z\right)$ is the transfer function of free space propagation, of which $i$ is the imaginary unit, $\lambda$ is the wavelength of beam, $(\nu_x,\nu_y)$ is the spatial frequency coordinates and $d_z$ is the beam transmission distance in the direction of $z$-axis.

In BRBL, the beam may transfer over the air between two planes which are not coaxis. Thus, as in Fig.~\ref{Fig.iteration}(a), the beam transmission over the air is supposed to be expressed as~\cite{delen1998free}:
\begin{equation}
\begin{aligned}
   \mathrm{T}_{\mathrm{fs}}(u)=&
    \mathcal{F}^{-1}\{\mathcal{F}\{u(x,y)\} 
     S\left(\nu_{x}, \nu_{y}, s_{x}, s_{y}\right)\\ 
    &\cdot H\left(\nu_{x}, \nu_{y}, d_z\right) \},
\end{aligned}
\end{equation}
where the function $S\left(\nu_{x}, \nu_{y}, s_{x}, s_{y}\right)$ is the phase delay that should be multiplied in the transmission wave as we calculate the Fourier transform in a shifted coordinate system, and is depicted as:
\begin{equation}
   S\left(\nu_{x}, \nu_{y}, s_{x}, s_{y}\right)=\exp \left[i 2 \pi\left(\nu_{x} s_{x}+\nu_{y} s_{y}\right)\right],
\end{equation}
where  $(s_x,s_y)$ illustrated in Fig.~\ref{Fig.iteration}(a) is the beam transmission distance in the direction of $x$-axis and $y$-axis, respectively.

Besides, the beam field will change after passing through the lens in the cat’s eye retroreflector, which can be expressed as~\cite{liu2021mobile,hodgson2005laser}:
\begin{equation}
    \mathrm{T}_{\mathrm{lens}}(u)=u(x,y) \exp \left[-i \frac{\pi}{\lambda f}\left(x^{2}+y^{2}\right)\right], x^{2}+y^{2} \leq r_l^{2}, 
\end{equation}

where $f$ and $r_l$ denote the focal length and the radius of the lens, respectively.

According to the self-reproducing mode theory, the beam field will change the amplitude and phase as propagating within the resonant cavity, while the distribution remains unchanged. Iterating the self-consistent equation, we can finally obtain the self-reproducing mode \cite{FoxLi}. The self-reproducing process is depicted as Fig.~\ref{Fig.iteration}(b). Starting from the mirror of the cat's eye retroreflector at the BS, i.e., M1, the self-consistent equation representing the round-trip beam transmission after passing through each element and the free space within the system can be formed as \cite{FFT,liu2021mobile}:
\begin{equation}
\label{equ:consistEq}
\begin{aligned}
\xi u =\mathrm{T}_{\mathrm{round }}(u), 
\end{aligned}
\end{equation}
where $\xi$ is the eigenvalue of this consistent equation, depicting the amplitude and phase change after a round-trip transmission; $u$ is the field distribution on the M1; $\mathrm{T}_{\mathrm{round}}$ is the transfer function of a round-trip transmission process. The round-trip transmission process can be resolved into forward and backward transfers. The round-trip is also depicted as the loop marked by red arrows in Fig.~\ref{Fig.RBS}, so the transfer function can be detailed as:
\begin{equation}
    \begin{aligned}
    &\mathrm{T}_{\mathrm{round }}(u)  =\mathrm{T}_{\mathrm{fwd}}\mathrm{T}_{\mathrm{bwd}},\\
&\mathrm{T}_{\mathrm{fwd}}(u)=\mathrm{T}_{\mathrm{bwd}}(u) = \mathrm{T}_{\mathrm{fs}}\mathrm{T}_{\mathrm{lens}}\mathrm{T}_{\mathrm{fs}}\mathrm{T}_{\mathrm{lens}}\mathrm{T}_{\mathrm{fs}}.
    \end{aligned}
\end{equation}
Moreover, the beam passes through every component of the resonant cavity, such as the cat's eye retroreflector in BS, the gain medium, the free space between two retroreflectors and the cat's eye retroreflector in target, twice in a round trip.

After adequate iterations, as depicted in Fig.~\ref{Fig.iteration}(b), the normalized beam field distribution will be identical to the one in last loop, only amplitude attenuation and phase lag happening~\cite{FoxLi}. Meanwhile, the self-reproducing mode can be described as:

\begin{equation}
\left|u_{q+1}\right|^2=|\xi|^2 \left|u_q\right|^2,
\end{equation}
where $q$ is the number of iterations, and $|\xi|^2$ is expressed as over-the-air transmission efficiency. 

Owing to there is no analytical solution to \eqref{equ:consistEq}~\cite{hodgson2005laser}, we calculate the approximate numerical solution through numerical simulation method. In addition, the general condition for iteration termination of simulation is that iteration stops when the standard deviation of the field distribution difference between $u_{q}$ and $u_{q+1}$ is less than $10^{-4}$ \cite{iteration}.

\subsection{Binocular Localization}


\begin{figure*}[!t]
	\centering
	\includegraphics[scale=0.75]{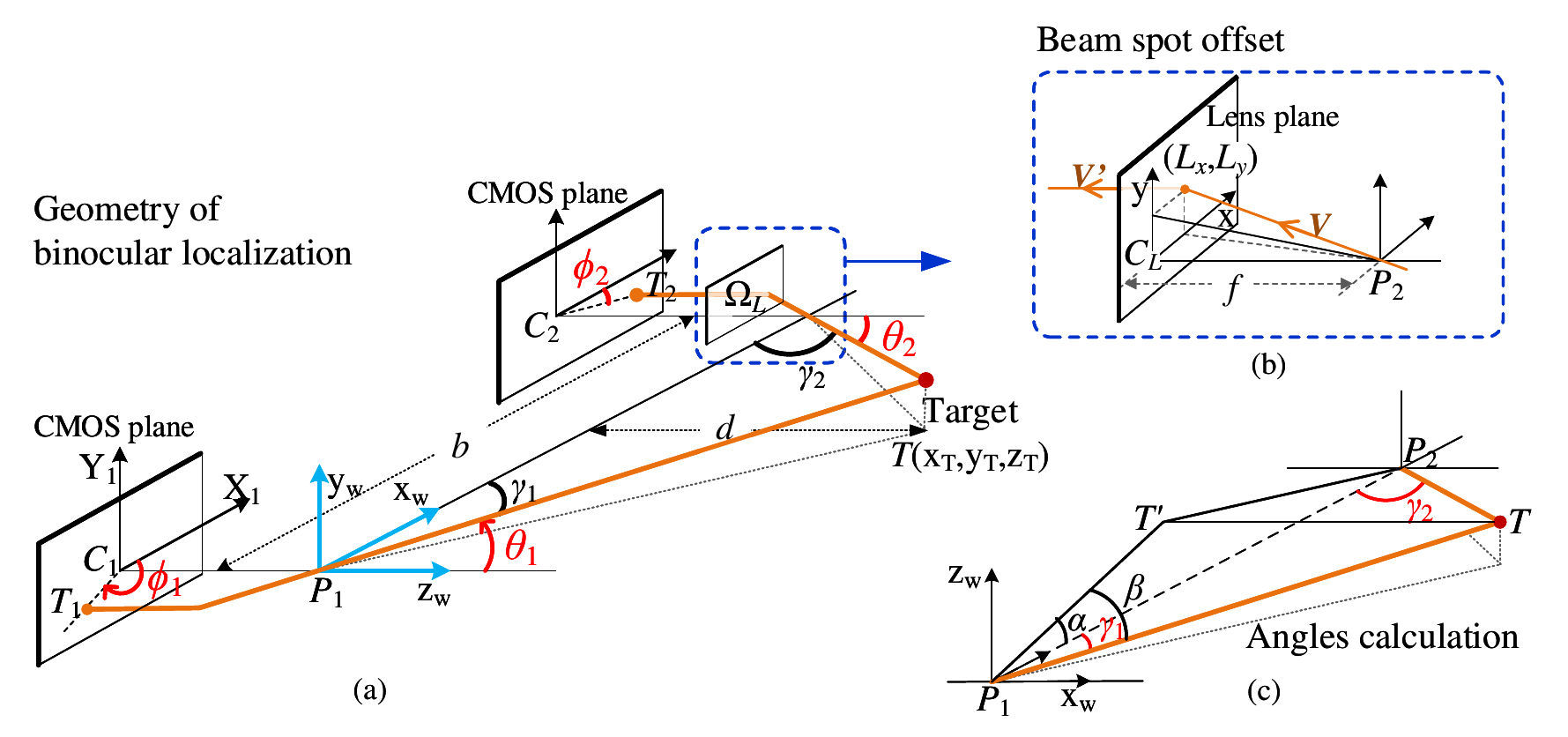}
	\caption{Geometry of binocular localization. (a) Beam incident angles and binocular disparity with specific spatial coordinate systems, (b) Transformation of the beam when it encounters a focal lens (taking TX2 as example), and (c) the angles calculation in 3-D localization derivation.}
	\label{Fig.biDisparity}
\end{figure*}

In this subsection, we will derive the incident angles of the resonant beams from the spot images exported from TXs, and then calculate the precise space coordinates of the target with errors laid aside.

\subsubsection{Beam incident angles}
It is easy to infer the resonant beam elevation angle and azimuth angle from a single CMOS image, which specify the polar angle $\theta$ and the azimuth angle $\phi$ of the target in a spherical coordinate system, respectively. Hence, we denote the incident angles as $\theta$ and $\phi$ in the following. To be specific, we adopt geometrical optics to describe how resonant beam propagates when it encounters the optical elements in TXs. 

We depict the geometry of the binocular localization in Fig.~\ref{Fig.biDisparity}. Supposing the target is a physical point $T$ in the 3-D space, the resonant beams emerge, then go through different elements, finally impinge on the CMOSs with the projected image points $T_1$ and $T_2$ in the left and right view images (i.e., CMOS plane), respectively. The incident angles are depicted as $\theta_1$, $\theta_2$, $\phi_1$ and $\phi_2$ (polar angles and azimuth angles of the beams striking CMOS1 and CMOS2, respectively). 

As shown in Fig.~\ref{Fig.biDisparity}(b), we take TX2 as an example to express the beam propagation process when it encounters a lens, and build a 2-D Cartesian coordinate system originating from the center of the lens $C_L$. The incident beam can be assumed as a ray $\boldsymbol{V}$ starting from $P_2$, which changes the transmission direction as $\boldsymbol{V}^\prime$ after encountering the focal lens. Owing to the straight propagation of resonant beam, the beam impinges on the lens plane $\Omega_{L}$ at $(L_x,L_y)$ with geometric significance depicted in Fig.~\ref{Fig.biDisparity}(b), and can be derived as
\begin{equation}
	\begin{aligned}
&	L_x=f\tan{\theta}\cos{\phi}, \\	
&	L_y=f\tan{\theta}\sin{\phi}, \\
	\end{aligned}
\end{equation}
where $f$ is the focal length of the cat's eye retroreflector. Furthermore, because the beams pass through the focus of the ideal lens, they will keep transferring in a straight line that parallels to $z$-axis, after going through the ideal lens. Thus, the spot's center coordinates on the CMOS plane equal to that on the lens plane, i.e., $(c_x,c_y) = (L_x,L_y)$.

Thus, the incident angles that we want estimate from a spot image, denoted as a vector $\boldsymbol{\Phi}=(\theta,\phi)^T$, have dependencies on  $(c_x,c_y)$ as: 
\begin{equation}
	\label{equ:incidentVec}
	\boldsymbol{\Phi}=\left[\begin{array}{c}
		\theta \\ \phi
	\end{array}\right]=
\left[\begin{array}{c}
\tan^{-1}\sqrt{c_x^2+c_y^2}/f\\
\tan^{-1}c_y/c_x
\end{array}\right].
\end{equation}

The beam outputs from the mirror at BS and then impinges on the CMOS, presenting the same distribution on both. From \eqref{equ:incidentVec}, the estimate of $\mathbf\Phi$ can be transformed to the estimate of the beam spot center on CMOS, i.e., $(c_x,c_y)$. Here, we adopt the centroid algorithm to calculate the beam center, as:

\begin{equation}
	\label{equ:centroidCal}
	\begin{array}{l} \hat c_{x}=\frac{\sum\limits_{(x, y) \in D_\mathrm{T}} x W(x, y)}{\sum\limits_{(x, y) \in D_\mathrm{T}} W(x, y)} ,	\\
		\\ \hat c_{y}=\frac{\sum\limits_{(x, y) \in D_\mathrm{T}} y W(x, y)}{\sum\limits_{(x, y) \in D_\mathrm{T}} W(x, y)},
	\end{array}
\end{equation}
where $(\hat{c}_{x},\hat{c}_{y})$ is the estimated center of the beam spot on CMOS; $D_\mathrm{T}$ is grey-scale image depicting the area illuminated by the beam on CMOS; $W(x,y)$ is the weight function of centroid algorithm, and is derived as:

\begin{equation}
	W(x, y)=\left\{\begin{array}{ll}I(x, y)-I_\mathrm{th}, & I(x, y) > I_\mathrm{th} \\ 0, & I(x, y)\leq I_\mathrm{th}\end{array},\right.
\end{equation}
where $I(x,y)$, depicted in Fig.~\ref{Fig.iteration}(c), is the intensity distribution of the beam on transverse $x$-$y$-plane, and $I_\mathrm{th}$ is the threshold intensity that we distinguish the beam spot intensity from the background.

So far, we have demonstrated that the offset of beam spot position on each CMOS presents function relation with beam incidence angles. Hence, as long as we get beam spot images from CMOSs, we could estimate the incident angles of resonant beams, i.e., $[\hat{\boldsymbol{\Phi}}_1,\hat{\boldsymbol{\Phi}}_2] = [\hat{\theta}_1,\hat{\phi}_1,\hat{\theta}_2,\hat{\phi}_2]$, in given reference frames.

\subsubsection{Target Localization}
Having obtained incident angles $[\hat{\boldsymbol{\Phi}}_1,\hat{\boldsymbol{\Phi}}_2]$, we can then calculate the 3-D coordinates of the target in any given reference frame.

First of all, we establish the world coordinate system (WCS) $(x_\mathrm{w},y_\mathrm{w},z_\mathrm{w})$ as shown in Fig.~\ref{Fig.biDisparity}(a). The WCS originates from the pupil, $P_1$, of TX1. Except when passing through the lens in retroreflector, the beam travels in straight lines. Thus, we depict the beam propagation process through a lens in the Fig.~\ref{Fig.biDisparity}(b). From the coordinates of point pair $(T_1,T_2)$, we can derive the target's 3-D location $T(x_\mathrm{T},y_\mathrm{T},z_\mathrm{T})$ in WCS. 

To simplify the derivative expression, we introduce $\gamma_1$, $\gamma_2$, as the angle between baseline and $TP_1$ and $TP_2$, respectively. As depicted in Fig.~\ref{Fig.biDisparity}(c), we denote the vertical projection of $T$ on the $xoy$ plane of WCS as $T^\prime$, and build a tetrahedron denoted as $P_1T^\prime TP_2$. According to the space angle formulas of tetrahedron, we can derive the cosine of $\gamma_1$ as:
\begin{equation}
	\label{equ:spaceAngle}
	\cos\gamma_1 = \cos\alpha\cos\beta+\sin\alpha\sin\beta\cos\angle_{T- P_1T^\prime-P_2},
\end{equation}
where $\angle_{T- P_1T^\prime-P_2}$ denotes the dihedral angle of the plane  $TP_1T^\prime$ and plane $P_2P_1T^\prime$, and apparently equals to $\pi/2$; $\alpha$ denotes the plane angle $\angle_{T^\prime P_1P_2}$, and equals to $\phi_{1}$, $\beta$ denotes the plane angle $\angle_{T^\prime P_1T}$, and equals to $\pi/2-\theta_{1}$. Thus, we rewrite \eqref{equ:spaceAngle} as:
\begin{equation}
	\cos{\gamma_1}=\sin{\theta_1} \cos{\phi_{1}}.
\end{equation}
Similarity, the cosine of $\gamma_2$ can be calculated as:
\begin{equation}
    \cos{\gamma_2}=-\sin{\theta_2} \cos{\phi_{2}}.
\end{equation}

With the baseline length $b$ known, the distance between target and each pupil can be calculated, denoted as $||T P_1||$ and $||T P_2||$. Specifically, $||T P_1||$ is calculated as:
\begin{equation}
    ||T P_1||=\frac{(\sin{\gamma_2})b}{\sin{(\gamma_1+\gamma_2)}}.
\end{equation}
Accordingly, with the binocular vision disparity obtained, the specific coordinate of target can be calculated as:
\begin{equation}
	\left[
	\begin{array}{c}
		x_\mathrm{T} \\ y_\mathrm{T}\\ z_\mathrm{T}
	\end{array}
	\right] = \left[
    \begin{array}{c}
   ||T P_1||\sin{\theta_1}\cos{\phi_1} \\
   ||T P_1||\sin{\theta_1}\sin{\phi_1} \\
   ||T P_1||\cos{\theta_1}
   \end{array}\right].
    \label{equ:Tcoord}
\end{equation}
Therefore, the estimate of target position $(\hat{x}_\mathrm{T},\hat{y}_\mathrm{T},\hat{z}_\mathrm{T})$ can be transformed from the estimate of incident angles from \eqref{equ:Tcoord}.

\subsection{Effective Localization Scope}
\begin{figure*}[!t]
	\centering
	\includegraphics[scale=0.8]{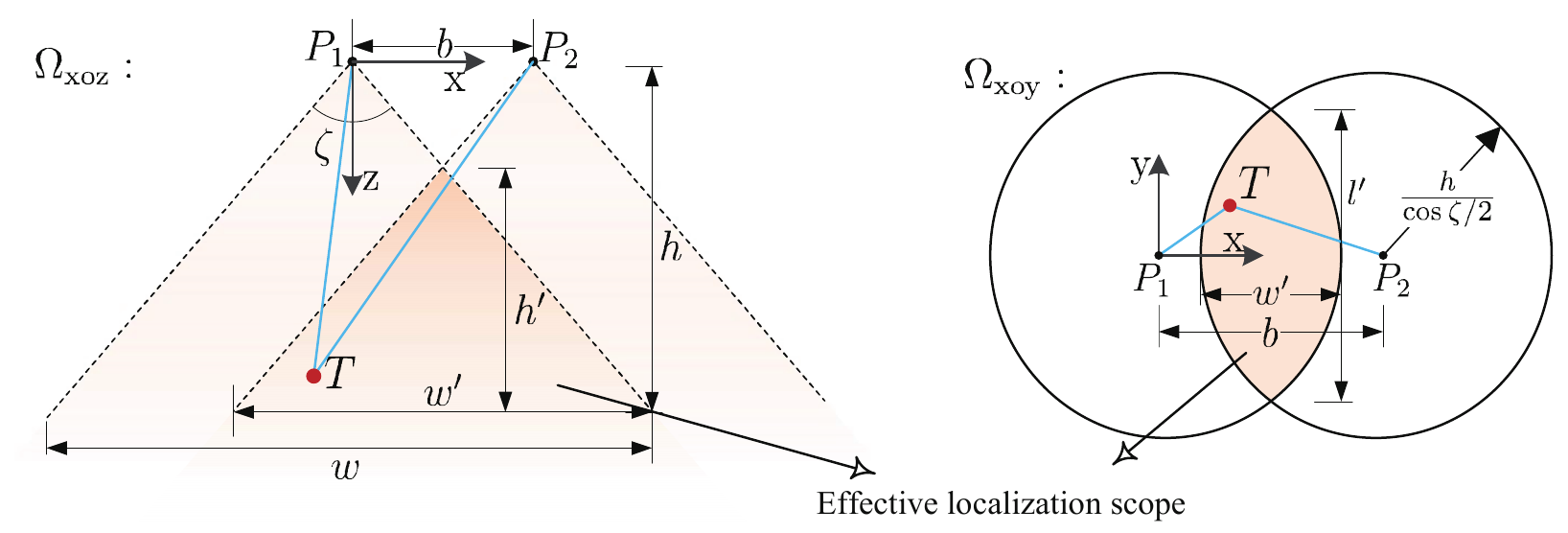}
	\caption{Effective localization scope of binocular vision illustrated in $\Omega_{\mathrm{xoz}}$ (i.e., plane $\mathrm{xoz}$) and $\Omega_{\mathrm{xoy}}$ (i.e., plane $\mathrm{xoy}$).}
	\label{Fig.biScope}
\end{figure*}

We use $\zeta$ to denote the FoV of monocular vision RBS, which depends on system parameters (cat's eye retroreflectors size, pump power, gain medium parameters, etc.) of the RBS. As in Fig.~\ref{Fig.biScope}, we depict the effective scope of binocular localization in the plane $xoz$ and plane $xoy$, respectively.

In the plane $xoz$, the maximum coverage width of single TX is
\begin{equation}
	\label{equ:w}
    w=2h\tan{\frac{\zeta}{2}},
\end{equation}
where $h$ is the height of the TX settled. If baseline length $b$ is given, the maximum boundaries of effective stereo scope can be expressed as:
\begin{equation}
    \left\{
    \begin{aligned}
    & w'=w-b \\
    & l'=\sqrt{\frac{4wh}{\cos{\zeta/2}}-w^2} \\
    & h'=\frac{(w-b)h}{w} 
    \end{aligned},
    \right.
    \label{equ:biscope}
\end{equation}
where $w'$ is the maximum FoV width both in plane $xoy$ and in plane $xoz$; $l'$ is the maximum FoV length in plane $xoy$ and $h'$ is the maximum FoV height in plane $xoz$.

\section{Analysis of System Errors}
\label{sec:erranalysis}

So far we have worked out the 3-D position of the target by binocular localization model with noises not considered. We now analyze the noises in the BRBL system and the impact on the localization. In detail, we first adopt the cyclic power model in the resonant cavity to obtain the beam power shining onto the CMOS. Moreover, we investigate various noise sources and the induced spatial errors correspondingly.

\subsection{Beam Power on CMOS}


Resonant beam experiences gain as well as loss in each transmission loop. The gain compensates for the loss if the steady state is reached. Then, the output beam power $P_{\mathrm{out}}$ can be obtained by the cyclic power model~\cite{koechner2013solid}:
\begin{equation}
    \begin{aligned}
    P_{\mathrm{out}}= 
    &A_\mathrm{b} I_\mathrm{s}\frac{ (1-R) \eta_{2}}{1-R \eta_2\eta_{3}+\sqrt{R \eta}\left(1/\left(\eta_{1}\eta_2 \eta_\mathrm{s}\right)-\eta_\mathrm{s}\right)} \\
    &\cdot\left[g_{0} \ell_\mathrm{g}-|\ln \sqrt{R \eta_\mathrm{s}^{2} \eta}|\right],\\
\end{aligned}
\end{equation}
where $A_\mathrm{b}$ is the overlap area of resonant beam and gain medium; $I_\mathrm{s}$ is the saturation intensity; $R=R_1R_2$ represents the coupled output factor; $\eta_1-\eta_4$ are the diffraction loss factors and $\eta=\eta_1\eta_2\eta_3\eta_4$ is the transmission factor for a loop; $\eta_\mathrm{s}$ is the loss factor of gain medium per transit; $g_0$ is the small signal coefficient and $\ell_\mathrm{g}$ is the length of gain medium. Moreover, $I_\mathrm{s}$ and $g_0$ can be yielded from a four-level system as: 
\begin{equation}
\label{equ:Is}
    I_\mathrm{s}=\frac{\hbar \nu}{\sigma \tau_\mathrm{f}}
\end{equation}
and
\begin{equation}
\label{equ:g0RateEqu}
    g_0=\sigma n_0 W_\mathrm{p}\tau_\mathrm{f},
\end{equation}
where $\hbar$ is Planck’s constant; $\nu$ is the frequency of the resonant beam; $\sigma$ is the stimulated emission cross section; $\tau_\mathrm{f}$ is the fluorescence decay time of the upper laser level in the gain medium; $n_0$ is the population density of ground level and $W_\mathrm{p}$ is the pump rate.


Besides, $W_\mathrm{p} n_0$ also gives the number of atoms transferred from the ground level to the upper laser level per unit time and volume, i.e., pump power density divided by photon energy, which is expressed as~\cite{koechner2013solid}:
\begin{equation}
\label{equ:WpN0}
    W_\mathrm{p} n_0=\frac{\eta_{\mathrm{excit}} P_{\mathrm{in}}}{\hbar \nu V_\mathrm{g}},
\end{equation}
where $V_\mathrm{g}$ is the volume of active gain medium, and $V_\mathrm{g}=A_\mathrm{b}\ell_\mathrm{g}$. Substituting \eqref{equ:Is} and \eqref{equ:WpN0} to \eqref{equ:g0RateEqu}, $g_0 \ell_\mathrm{g}$ can be depicted as:
\begin{equation}
    g_0 \ell_\mathrm{g} =\frac{\eta_{\mathrm{excit}} P_{\mathrm{in}}\ell_\mathrm{g}}{I_\mathrm{s} V_\mathrm{g}}=\frac{\eta_{\mathrm{excit}} P_{\mathrm{in}}}{I_\mathrm{s}A_\mathrm{b}},
\end{equation}
where $P_{\mathrm{in}}$ is the pump input power, $\eta_{\rm{excit}}$ is the excitation efficiency of the gain medium.

In BRBL, an attenuation piece is adopted in front of the CMOS in each TX, so that the incident beam intensity can satisfy the restriction of CMOS full-well capacity. Part of the resonant beam leaked from the coupling output mirror M1, experiences last diffraction transmission and finally arrives at the CMOS with the power $ P_{\mathrm{C}}$: 

\begin{equation}
    P_{\mathrm{C}}=\rho \eta_\mathrm{c} P_{\mathrm{out}}, 
\end{equation}
where $\rho$ is the attenuation factor of the attenuation tablets, $\eta_\mathrm{c} $ is the loss factor of the transmission process from the mirror M1 to the CMOS.

\subsection{Linear Signal Model for Imaging}
\begin{figure}[!t]
    \centering
     \includegraphics[scale=0.8]{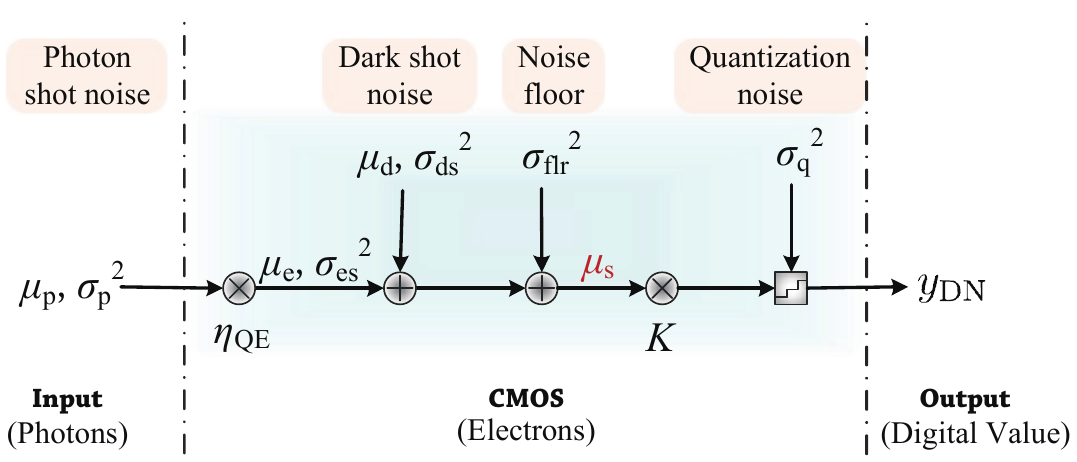}
    \caption{Linear signal model and noise model in the imaging process of the CMOS~\cite{european2010standard}.}
    \label{Fig.noiseModel}
\end{figure}
The quality of the spot on CMOS, which is limited by various system noises, directly influences the location estimation performance. Resonant beam illuminates the CMOS and stimulates photo-induced electrons which are collected by pixels. We depict the linear signal model and the noise model in the imaging process in Fig.~\ref{Fig.noiseModel}. The response characteristic curve of CMOS is generally linear as in Fig.~\ref{Fig.noiseModel} in its dynamic range~\cite{european2010standard}:
\begin{equation}
\label{equ:DigitalNo.}
    y_{\mathrm{DN}} = K(\mu_\mathrm{e}+\mu_\mathrm{d}),
\end{equation}
where $y_{\mathrm{DN}}$ is the output digital signal by an analog digital converter (ADC) in CMOS; $K$ is the sensitivity of the CMOS with units $\mathrm{DN/e-}$, i.e., digits per electrons; $\mu_\mathrm{e}$ is the number of photo-induced electrons accumulated during the exposure time $t_{\mathrm{exp}}$; $\mu_\mathrm{d}$ is the mean number of electrons present without beam, also known as dark current. Hence, the signal here can be described in charge units as $\mu_\mathrm{s}$:
\begin{equation}
\label{equ:mu_s}
    \mu_\mathrm{s}=\eta_{\mathrm{QE}}\mu_\mathrm{p}+\mu_\mathrm{d},
\end{equation}
where $\mu_\mathrm{e} = \eta_{\mathrm{QE}}\mu_\mathrm{p}$; $\eta_{\mathrm{QE}}$ is quantum efficiency at wavelength $\lambda$; $\mu_\mathrm{p}$ is the number of photons that hit a pixel, representing the photon irradiance. 

Moreover, introducing the photoelectric conversion, we shall relate the resonant beam power to the CMOS output signal:
\begin{equation}
\mu_\mathrm{p}=\frac{A I t_{\mathrm{exp}}}{\hbar \nu}=\frac{A I t_{\mathrm{exp}}}{\hbar c/\lambda},
\label{equ:mu_p}
\end{equation}
where $A$ is the area of a pixel; $I$ is the irradiance that resonant beam illuminating on the CMOS; $t_{\mathrm{exp}}$ is the exposure time and $c$ is the speed of light.

With \eqref{equ:mu_s} and \eqref{equ:mu_p}, \eqref{equ:DigitalNo.} can be expressed as a linear relationship between the mean grey value $y_{\mathrm{DN}}$ and the number of photons  irradiated from the beam during the exposure time. Thus, the linear signal model can be expressed as:
\begin{equation}
    y_{\mathrm{DN}} =K\eta_{\mathrm{QE}}\frac{A I t_{\mathrm{exp}}}{\hbar c/\lambda}+K\mu_\mathrm{d}.
\end{equation}

\subsection{Noise and Error Analysis}

The noises deteriorating images include temporal noise and fixed pattern noise (FPN). Temporal noise fluctuates over time, while FPN is fixed in space and can be removed by signal processing in principle~\cite{nakamura2017image}. Hence, We regard the FPN as background in~\eqref{equ:centroidCal} and eliminate it in the centroid calculation.

Fig.~\ref{Fig.noiseModel} also shows the noises generated in the whole process, from photons to electrons then to digital value. We analyse various temporal noises including shot noise, noise floor, quantization noise, etc. Since these noise sources are uncorrelated, we can express the total noise power by adding up these noises' variance as:
\begin{equation}
\label{equ:totNoise}
	\sigma_\mathrm{tot}^2=K^2(\sigma_{\mathrm{sh}}^{2}+\sigma_{\mathrm{flr}}^{2})+\sigma_{\mathrm{q}}^{2},
\end{equation}
where $\sigma_\mathrm{tot}^2$ is the total temporal noise variance of the signal $y_{\mathrm{DN}}$; $\sigma_{\mathrm{flr}}^2$ is the variance of noise floor containing read noise and amplifier noise in circuit device; $\sigma_{\mathrm{q}}^2$ is the variance of noise in ADC; $\sigma_{\mathrm{sh}}^2$ is the variance of shot noise including the dark current shot noise $\sigma_{\mathrm{ds}}^2$ and the photon-electron shot noise $\sigma_{\mathrm{es}}^2$, and $\sigma_{\mathrm{sh}}^2=\sigma_{\mathrm{ds}}^2+\sigma_{\mathrm{es}}^2$ ~\cite{nakamura2017image}.




Moreover, the possibility of shot noise complies with Poisson distribution, with the noise variance of the fluctuations equivalent to the mean number of accumulated electrons. Thus we can obtain the photon-electron shot noise and the dark current shot noise as~\cite{nakamura2017image}:
\begin{equation}
\begin{aligned}
\label{equ:shotNoise}
    &\sigma_{\mathrm{es}}^2=\mu_\mathrm{e}=\eta_{\mathrm{QE}}\frac{A I t_{\mathrm{exp}}}{\hbar c/\lambda}, \\
    &\sigma_{\mathrm{ds}}^{2}=\mu_\mathrm{d}=\frac{J_{\mathrm{dark}} A t_{\mathrm{exp}}}{q},
\end{aligned}
\end{equation}
where $\sigma_{\mathrm{ds}}^2$ depicts variance of the mean dark current shot noise relying on the dark current $J_{\mathrm{dark}}$ under specific operating temperature and $J_{\mathrm{dark}}$ is often provided on CMOS spec. sheet; $q$ is electric charge. With \eqref{equ:totNoise} and \eqref{equ:shotNoise}, the total noise can be derived as:
\begin{equation}
    \sigma^2_\mathrm{tot}=K^2(\mu_\mathrm{e}+\mu_\mathrm{d})+K^2\sigma_{\mathrm{flr}}^{2}+\sigma_{\mathrm{q}}^{2}.
\end{equation}

In addition, the sensitivity $K$, the noise floor $\sigma_{\mathrm{flr}}^2$, the quantization noise $\sigma_{\mathrm{q}}^{2}$ are also in general provided on spec. sheet. The quantization noise has a tiny effect on the image and can be integrated into the noise floor~\cite{european2010standard}. Thus, the overall system gains $K$ is cancelled out in the signal-noise ratio (SNR) of BRBL, which can be expressed as:
\begin{equation}
\label{equ:snrCal}
    \mathrm{SNR_{dB}} =20\log_{10}\left(\frac{\eta_{\mathrm{QE}}\mu_\mathrm{p}}{\sqrt{\eta_{\mathrm{QE}} \mu_\mathrm{p}+\mu_\mathrm{d}+\sigma^2_{\mathrm{flr}}}}\right).
\end{equation}

As demonstrated above, we can only estimate the spot center with noises mixed by the centroid calculation in \eqref{equ:centroidCal}. Due to the received signal interfered by various noises, the spatial error of centroid calculation is inevitable, and can be expressed by the root mean square error (RMSE) as:
	\begin{equation}
\label{equ:errCal}
    \mathrm{RMSE_{spot}}=\sqrt{(\hat c_x -c_x)^2+(\hat c_y -c_y)^2},
\end{equation}
where $(\hat{c}_x,\hat{c}_y)$ is the estimated center of beam spot on CMOS, and $(c_x,c_y)$ is the precise coordinates of the target mapping point on CMOS. 

Accordingly, having obtained the estimated center of beam spot $(\hat{c}_x,\hat{c}_y)$ on each CMOS, we can figure out the beam incident angles as $\hat\theta$ and $\hat\phi$. From the binocular localization model above, the estimated location of target $\hat T(\hat{x}_\mathrm{T},\hat{y}_\mathrm{T},\hat{z}_\mathrm{T})$ is finally calculated, of which the errors are due to the noises in the system. The localization errors can be depicted as:
\begin{equation}
\label{equ:tgtErrCal}
    \mathrm{RMSE_{tgt}}=\sqrt{||\hat T -T||^2}.
\end{equation}

\section{Numerical Analysis and Results}
\label{sec:simul}
In this section, we evaluate the target position estimation accuracy numerically and give a discussion on some open issues. Firstly, by simulating the beam transmission process in resonant cavity and the conversion process in CMOS, we depict the beam spot positions and RMSE of spot center on the CMOS plane concerning different target positions. Then, we compare the SNR of each beam spot under various power conditions to illustrate the influences of input power and noises. At last, we calculate the 3-D localization errors of BRBL with the target position changing in different directions and demonstrate that the localization accuracy can be generally at the centimeter level.

\subsection{Parameters}
In this subsection, we exhibit the parameters in numerical analysis for ``Beam Field Transmission Simulation'', ``Cyclic Power Model'' and ``Linear Signal Model''.

\begin{table}[!htbp]
\newcommand{\tabincell}[2]{\begin{tabular}{@{}#1@{}}#2\end{tabular}}
\centering
\caption{~Parameters for beam field transmission Simulation~\cite{hodgson2005laser}}
\vspace{.7em}
\begin{tabular}{ccc}
\hline
\textbf{Symbol}&\textbf{Parameter}&\textbf{Value}\\
\hline
$r_m/r_l$&\text{Cat's eye radius}&$2.5$mm\\
$r_g$&\text{Gain medium radius}&$2.5$mm\\
$f$&\text{Focal length of L1/L2}&$10$mm\\
$l$&\text{Distance between L1/L2 and M1/M2}&$10.01$mm\\
$l_2$&\text{Distance between M1 and CMOS}&$10$mm\\
$\lambda$ &\text{Resonant beam wavelength}& $1064$nm\\
$S_\mathrm{N}$&\text{FFT sampling number}&$8192$\\
$G$ &\text{Computation window expansion factor}& $3$\\
\hline
\label{Tab.fieldTrans}
\end{tabular}
\end{table}

\begin{table}[!htbp]
\newcommand{\tabincell}[2]{\begin{tabular}{@{}#1@{}}#2\end{tabular}}
\centering
\caption{~Parameters for Cyclic Power Model}
\begin{tabular}{ccc}
\hline
\textbf{Symbol}&\textbf{Parameter}&\textbf{Value}\\
\hline
$R$&\text{Output coupler reflectivity}&$99.9\%$\\
$I_\mathrm{s}$&\text{Medium saturated intensity}&$1.26\times10^{7}$ W/m$^2$\\
$\eta_\mathrm{s}$&\text{Loss factor in medium}&$0.99$\\
$\eta_{\mathrm{excit}}$&\text{Excitation efficiency}&$0.72$\\
$R_1$&\text{Reflectivity of M1 }&$99.9\%$\\
$\eta_\mathrm{c}$ &\text{Loss factor between CMOS and M1}& $0.99$\\
$\rho$ &\text{Attenuation factor}& $10^{-4}$\\
\hline
\label{Tab.cyclicPara}
\end{tabular}
\end{table}

\begin{table}[!htbp]
\newcommand{\tabincell}[2]{\begin{tabular}{@{}#1@{}}#2\end{tabular}}
\centering
\caption{~Parameters for Linear Signal Model}
\begin{tabular}{ccc}
\hline
\textbf{Symbol}&\textbf{Parameter}&\textbf{Value}\\
\hline
$\hbar$ &\text{Planck's constant}&$6.6260755\times 10^{-34}$ Js\\
$c$ &\text{Speed of light}& $2.99792458\times10^8$ m/s\\
$k$&\text{Boltzmann constant} & $1.38064852\times 10^{-23}$ J/K \\
$q$&\text{Electric charge unit} & $1.602176634\times10^{-19}$ C \\
$p_\mathrm{s}$ &\text{CMOS pixel size}& $1.8\mu$m\\
$N_\mathrm{s}$ &\text{CMOS pixel number}& $8192$\\
$t_{\mathrm{exp}}$ &\text{CMOS exposure time} &  $50\mu$s\\
$\eta_{\mathrm{QE}}$ &\text{Quantum efficiency} &  $80\%$\\
$J_{\mathrm{dark}}$ & \text{Dark current} & $10$ nA/cm$^2$\\
$\sigma^2_{\mathrm{flr}}$ & \text{Noise floor} & $1 \mathrm{e^-}$(rms)\\
\hline
\label{Tab.signalModel}
\end{tabular}
\end{table}

The parameters of field transmission simulation in the resonant cavity are listed in Table~\ref{Tab.fieldTrans}. The radius of each component in the cat's eye retroreflector, the mirror M1(M2) and the lens L1(L2), denoted as $r_m$ and $r_l$, are equal to that of the gain medium, $r_g$. Besides, the distance between CMOS and mirror in cat's eye is equal to the focal length, i.e., $l_2=f$. We have also displayed the parameters for FFT-based calculation and numerical iteration. Therein, $S_\mathrm{N}$ is the sampling number, which is generally a power of $2$ and should satisfy the precision requirement.

For cyclic power model, considering the CMOS is limited by full-well capacity, we set the output coupled reflectivity $R$ of the mirrors in TX close to $1$, so that beam power impinging on the CMOS is low. For the same purpose, the attenuation piece has an attenuation factor $\rho$, which can reduce the laser intensity outside of the cavity by several orders of magnitudes. $I_\mathrm{s}$, $\eta_\mathrm{s}$ and $\eta_{\rm{excit}}$ are determined by the physical properties of the gain medium. The detailed parameters are in Table~\ref{Tab.cyclicPara}.

We assume that the CMOS can be described by a linear signal model. In Table~\ref{Tab.signalModel}, the parameters that characterize the CMOSs are listed, with most values referring to~\cite{european2010standard,chi2011noise}. Therein, the units of the noise floor is in root-mean-square (rms) electrons, and the mean dark current is $10\mathrm{nA/cm}^2$ assuming that the temperature is $300\mathrm{K}$. 

\subsection{Spot Deviation on CMOS}
\begin{figure}[!t]
	\centering
	\includegraphics[scale=0.58]{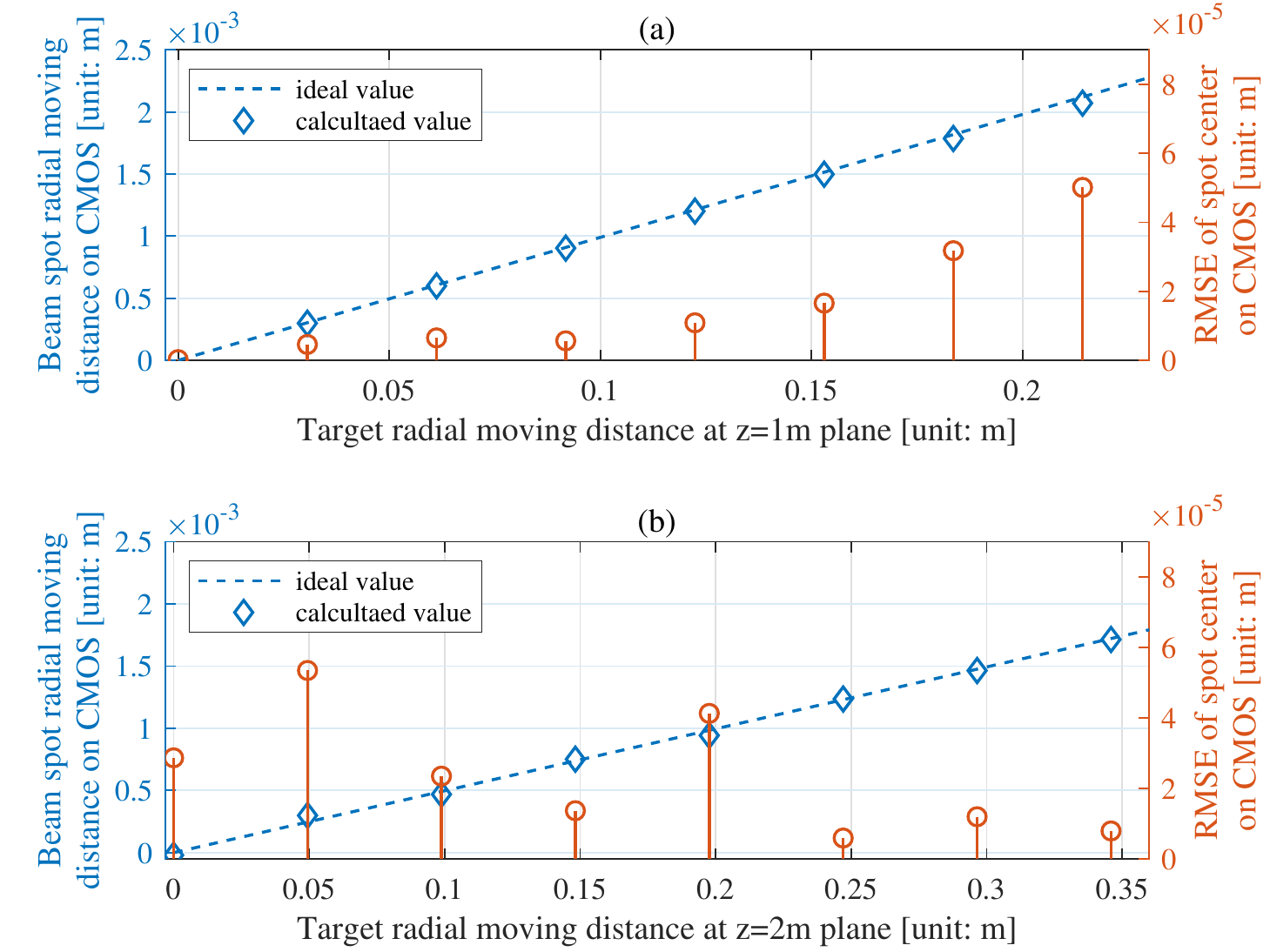}
	\caption{Beam spot positions on CMOS and RMSE of the spot center (taking CMOS1 in TX1 as example) when target moves along an arbitrary radial direction (a) at $z=1\mathrm{m}$ plane and (b) at $z=2\mathrm{m}$ plane, respectively. Meanwhile, the pump power $P_{\mathrm{in}}$ is $200\mathrm{W}$.}
	\label{Fig.SpotErr}
\end{figure}

We calculate the spot center using the centroid algorithm in \eqref{equ:centroidCal} and compare the spot positions with the ideal values, which validates the good collimation and high accuracy of the BRBL. In Fig.~\ref{Fig.SpotErr}, we depict the calculated spot center when the target moves different distances along an arbitrary radial direction at $z=1\mathrm{m}$ plane and at $z=2\mathrm{m}$ plane (i.e., the $z$-axis projection $d$ is $1\mathrm{m}$ and $2\mathrm{m}$), respectively. Moreover, the target's maximum radial moving distance differs, i.e., $24.48\mathrm{cm}$ at $z=1\mathrm{m}$ plane and $39.53\mathrm{cm}$  at $z=2\mathrm{m}$ plane. However, when the target is at the maximum radial distance, the intensity of beam spot is too low to be detected and the calculation of center is beyond reach. At this time, the angle between beam and $z$-axis is half of the monocular FoV, $\zeta/2$.
	

Due to the straight line transmission, the more deflected the target is from the perpendicular direction, the larger offset the beam spot gets. The ideal offset values of spot position are illustrated as the dashed lines in Fig.~\ref{Fig.SpotErr}. However, as we obtain the beam spot images by numerical analysis of beam transmission and the simulation of CMOS imaging, the calculated beam spot center gets inevitably perturbed. Therefore, the diamond points in Fig.~\ref{Fig.SpotErr} represent the calculated spot center when the radial moving distance of target changes at $z=1\mathrm{m}$ plane and at $z=2\mathrm{m}$ plane, respectively, with pump power $P_\mathrm{in}$ equal to $200\mathrm{W}$.

According to \eqref{equ:errCal}, we depict the deviations of the spot positions, i.e, $\mathrm{RMSE_{spot}}$, against the ideal values in Fig.~\ref{Fig.SpotErr}, which is shown as the round dots and solid lines subject to the right axis. In regards to the error magnitude, when the target is at $z=1\mathrm{m}$ plane, the deviations shown in Fig.~\ref{Fig.SpotErr}(a) are almost a couple of microns and get larger with the increase of the target radial displacement, complying with intuitive comprehension. As in Fig.~\ref{Fig.SpotErr}(b), when the target is at $z=2\mathrm{m}$ plane, the trend of the deviations is less apparent in that the increase of cavity length brings higher requirements on the numerical calculation accuracy. However, influenced by various noises, such as shot noise, read noise, etc., the spot deviations are still within several microns, i.e., micron class resolution.

\subsection{Influence of Input Power and Noise}

\begin{figure}[t]
    \centering
     \includegraphics[scale=0.62]{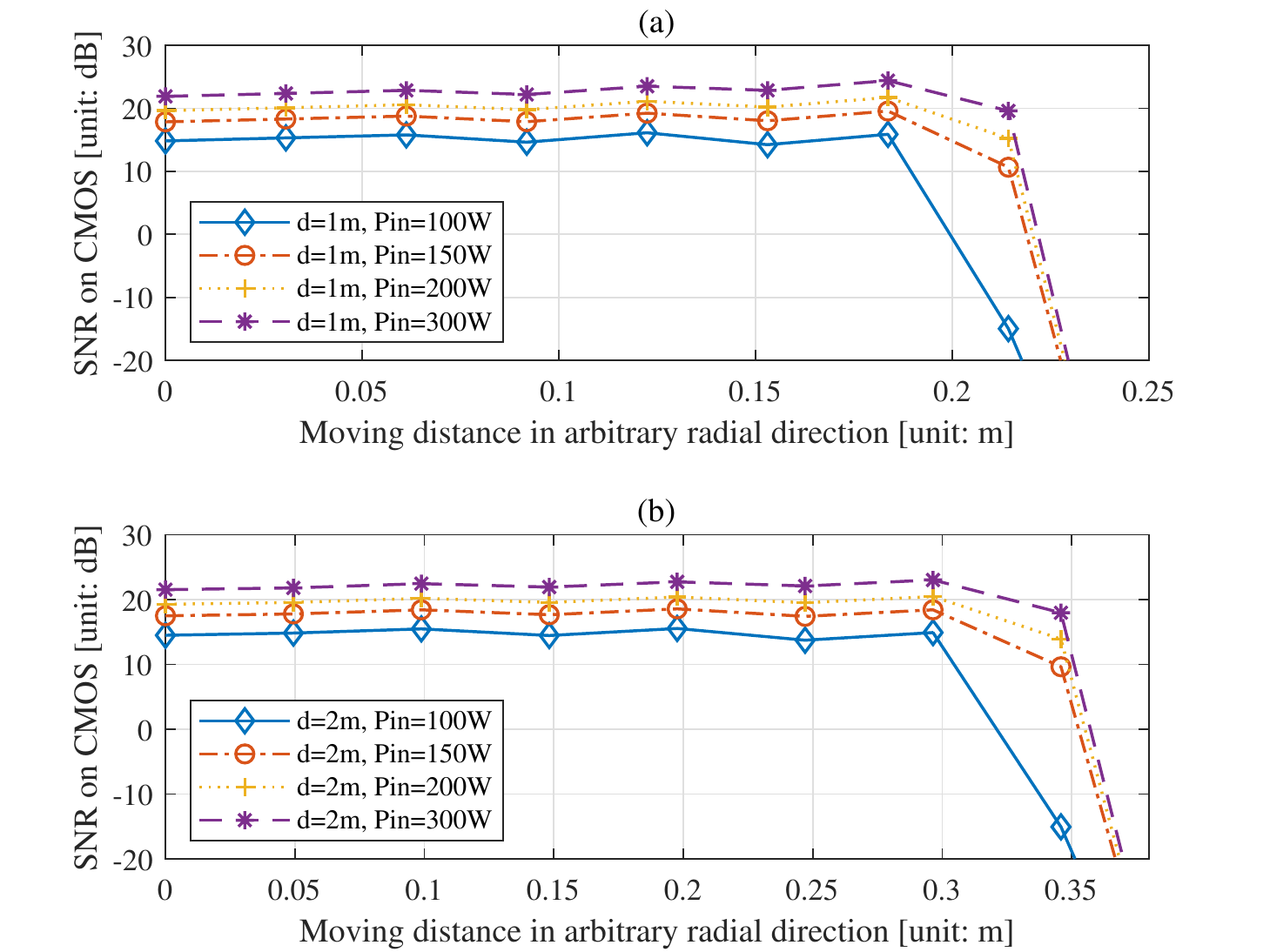}
    \caption{SNR (taking CMOS1 in TX1 as example) under various pump power situations when target's $z$-axis projection $d$ is (a) $1\mathrm{m}$ and (b) $2\mathrm{m}$. Meanwhile, pump power $P_{\mathrm{in}}$ is $100\mathrm{W}$, $150\mathrm{W}$, $200\mathrm{W}$ and $300\mathrm{W}$, respectively, and size of each CMOS is $2.5\times2.5 \mathrm{mm}^2$.}
    \label{Fig.SNRdB}
\end{figure}

Because of the symmetrical characteristic of binocular localization, we here show the SNR of the imaging process on CMOS1 in TX1. We depict the SNR at different pump power as the target position changes from the perpendicular direction to the FoV boundary with $d=1\mathrm{m}$ in Fig.~\ref{Fig.SNRdB}(a) and with $d=2\mathrm{m}$ in Fig.~\ref{Fig.SNRdB}(b), respectively.

The SNR only slightly fluctuates around a mean value, as the transverse distance between the target and $z$-axis increases within FoV, shown in Fig.~\ref{Fig.SNRdB}. That suggests the resonant beam is highly energy-concentrated once the stable resonator is established. When the target gets close to FoV boundary, i.e., about $21.48\mathrm{cm}$ with $d=1\mathrm{m}$ and about $34.59\mathrm{cm}$ with $d=2\mathrm{m}$, the SNR shows a significant decline. 

We also depict the SNR under different noises where input power $P_{\mathrm{in}}$ is $100\mathrm{W}$, $150\mathrm{W}$, $200\mathrm{W}$ and $300\mathrm{W}$, respectively, in Fig.~\ref{Fig.SNRdB}. Because the total temporal noises including shot noise, increase with the increase of signal intensity, the SNR on CMOS invariably fluctuates around $20\mathrm{dB}$ with $d=1\mathrm{m}$ and $d=2\mathrm{m}$. With the growth of the input power $P_{\mathrm{in}}$, the SNR increases slightly, as the output power on CMOS $P_\mathrm{C}$ is restricted in a narrow range for CMOS to harvest photons.


\subsection{Localization Accuracy}

\begin{figure*} \centering    
	\subfigure[Target moves along the $x$-axis in WCS with $d=1\mathrm{m}$.] { 
		\label{Fig.LocatErr_x_z1}  
		\includegraphics[width=0.95\columnwidth]{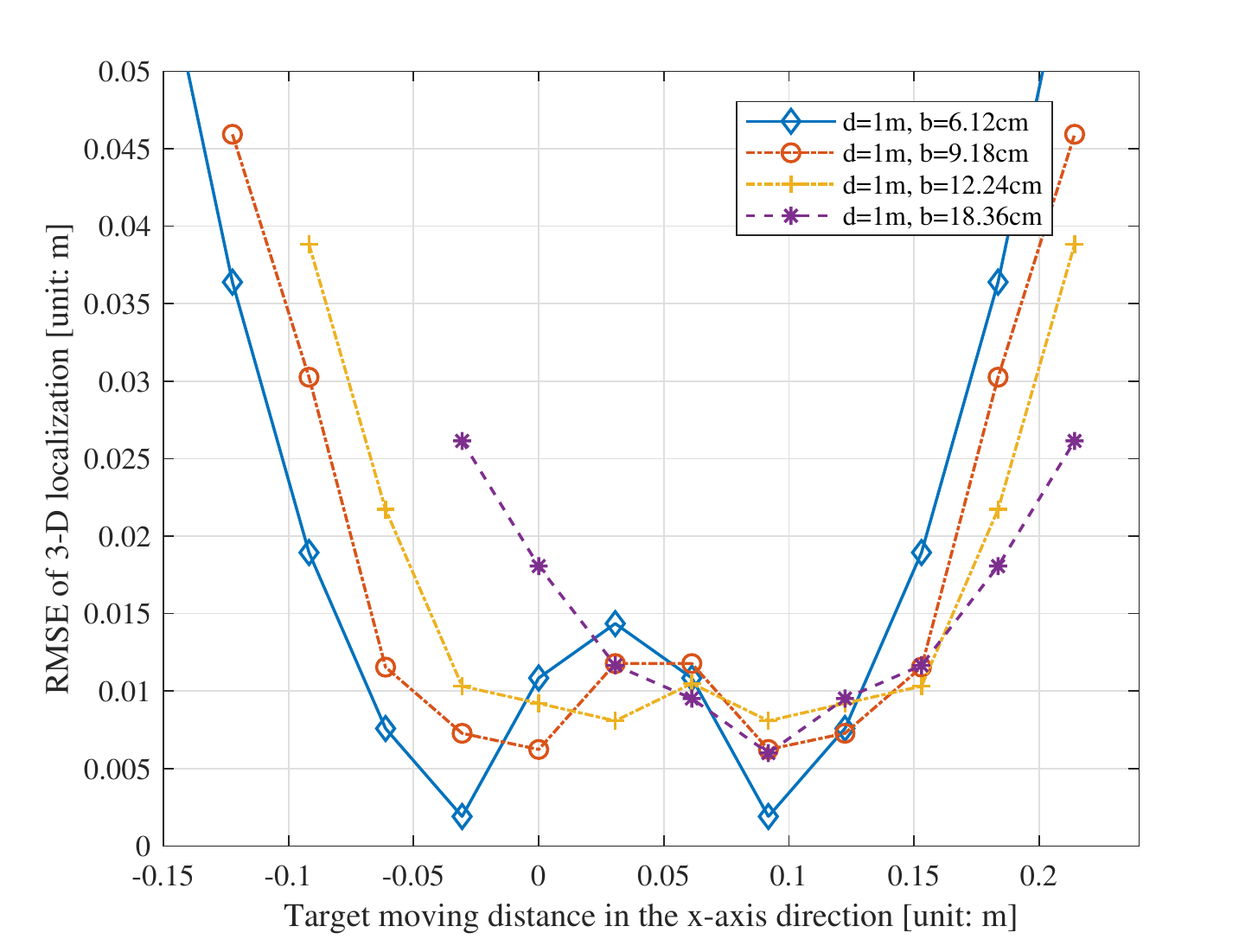}  
	}  
	\subfigure[Target moves along the $y$-axis in WCS with $d=1\mathrm{m}$.] { 
		\label{Fig.LocatErr_y_z1}     
		\includegraphics[width=0.95\columnwidth]{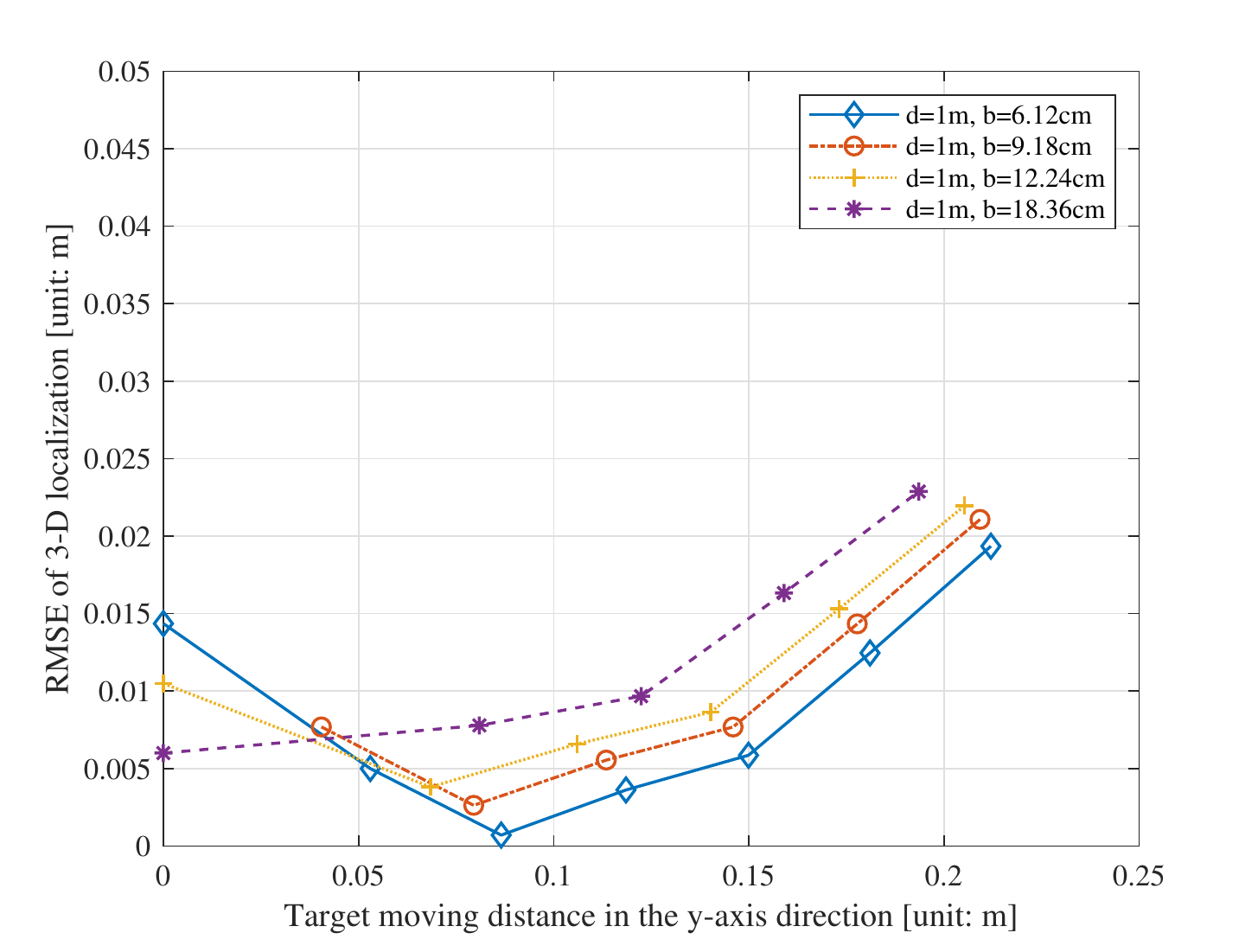}  
	} 
	\subfigure[Target moves along the $x$-axis in WCS with $d=2\mathrm{m}$.] { 
		\label{Fig.LocatErr_x_z2}     
		\includegraphics[width=0.95\columnwidth]{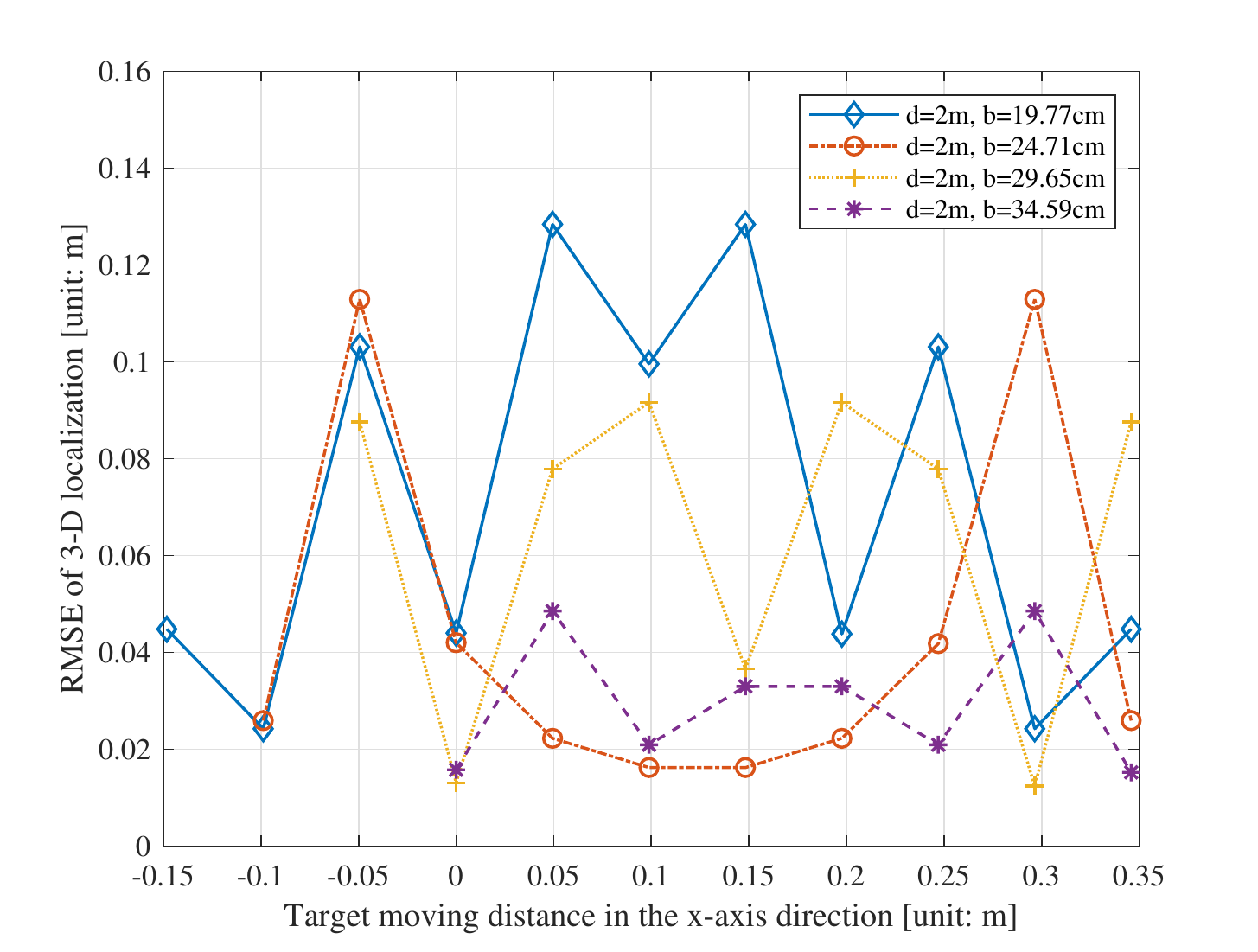}  
	}  
	\subfigure[Target moves along the $y$-axis in WCS with $d=2\mathrm{m}$.] { 
		\label{Fig.LocatErr_y_z2}    
		\includegraphics[width=0.95\columnwidth]{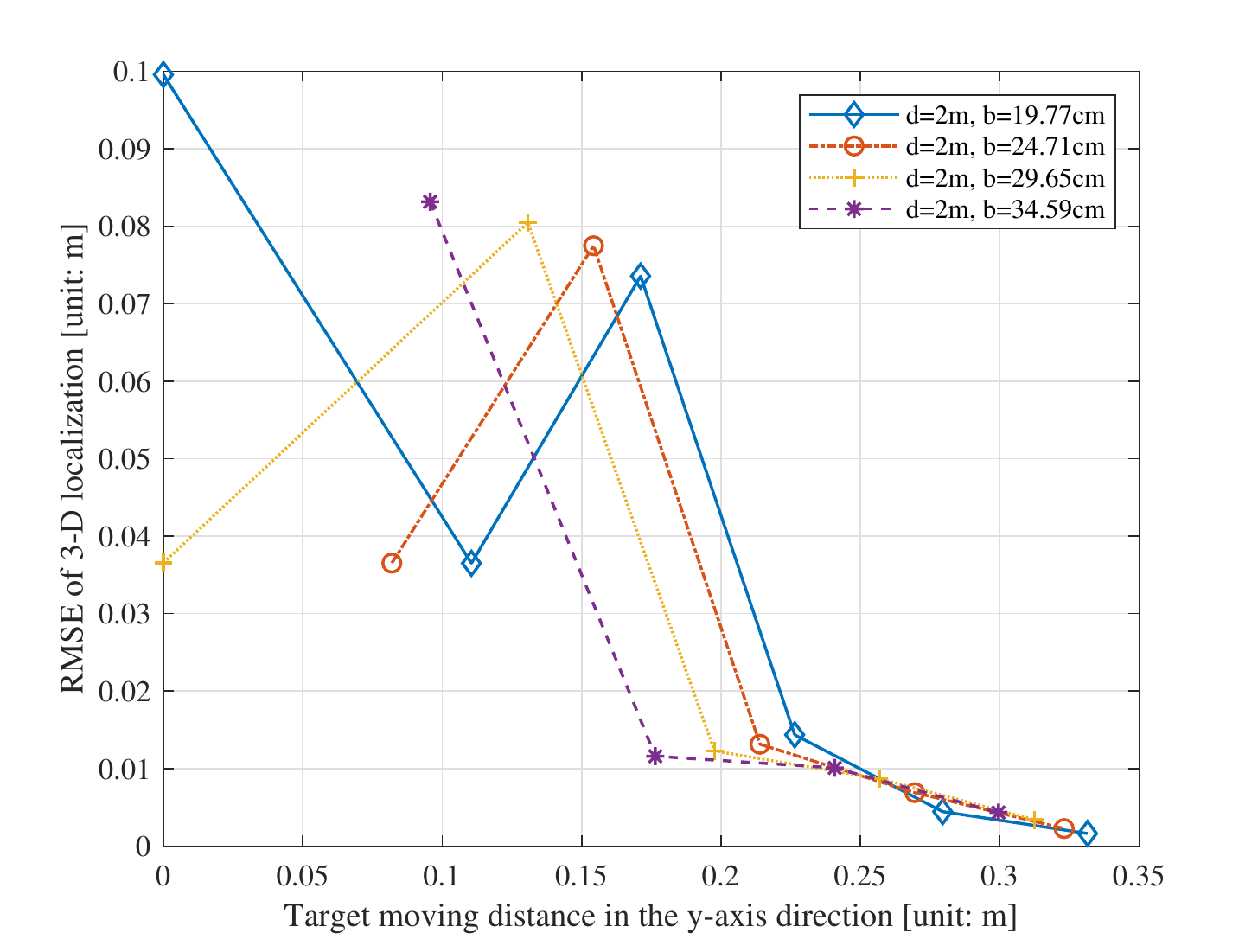}  
	} 
	\caption{RMSE of 3-D localization with different target positions.}     
	\label{fig:RMSE}     
\end{figure*}


To facilitate the computation, we apply the simulated spot images of TX1 in the 3-D localization calculation and obtain the estimated 3-D coordinates of the target $\hat T(\hat{x}_\mathrm{T},\hat{y}_\mathrm{T},\hat{z}_\mathrm{T})$. The localization accuracy is depicted as Fig.~\ref{fig:RMSE}.

We depict the deviations of estimated 3-D position against the precise coordinates of target $T(x_\mathrm{T},y_\mathrm{T},z_\mathrm{T})$ as $\mathrm{RMSE_{tgt}}$ in Fig.~9(a),(b) with $d=1\mathrm{m}$, and the deviations for $d=2\mathrm{m}$ are in Fig.~9(c),(d). To reduce the contingency of analysis on 3-D localization accuracy, we calculate $\mathrm{RMSE_{tgt}}$, when the target is at different positions, i.e., different positions with $(y_\mathrm{T},z_\mathrm{T})=(0,d)$, and different positions with $(x_\mathrm{T},z_\mathrm{T})=(b/2,d)$. The localization results are analyzed with $200\mathrm{W}$ input power and various baseline under each $d$.

From the results illustrated above, the deviations with $d=2\mathrm{m}$ generally show wider fluctuations than those with $d=1\mathrm{m}$, due to the erratic results brought about by the increase of resonant cavity length. Specifically, if the target is moving along the $x$-axis direction and is within the effective scope of BRBL at a certain baseline, the localization errors for $d=1\mathrm{m}$ are no more than $5\mathrm{cm}$. If the target moves along the $y$-axis direction, the localization error gets smaller, i.e., less than $2.5\mathrm{cm}$. While the localization errors can even exceed $10\mathrm{cm}$ for several points in the target's movement along the $x$-axis direction when $d$ is $2\mathrm{m}$. The localization errors in the target's movement along the $y$-axis with $d=2\mathrm{m}$ are likewise smaller than those along the $x$-axis direction, which are mostly less than $5\mathrm{cm}$ except for a few points.

%

In regards to the baseline, it is obvious that the choice of baseline is quite critical for high-accuracy localization. Hence, we also depict localization errors with various baselines in above figures, namely $6.12\mathrm{cm}$, $9.18\mathrm{cm}$, $12.24\mathrm{cm}$ and $18.36\mathrm{cm}$ for $d=1\mathrm{m}$, $19.77\mathrm{cm}$, $24.71\mathrm{cm}$, $29.65\mathrm{cm}$ and $34.59\mathrm{cm}$ for $d=2\mathrm{m}$. From the Fig.~\ref{Fig.LocatErr_x_z1} and Fig.~\ref{Fig.LocatErr_x_z2}, the localization errors with different baselines are invariably symmetric about the $b/2$ when the target's position changes along the $x$-axis. As in Fig.~\ref{Fig.LocatErr_y_z1} and Fig.~\ref{Fig.LocatErr_y_z2}, we depict the localization errors with various baselines when the target moves along the $y$-axis direction starting from $y=0\mathrm{cm}$. It should be noted that the errors are quite large with the target location close to both of the two TXs, because the closer the target is getting between these two TXs, the greater the influence of spot location errors is.


Besides, as demonstrated in~\eqref{equ:biscope}, the effective scope of BRBL is subject to the baseline because of the limited FoV of monocular RBS. The larger the baseline is, the smaller the effective scope of BRBL is. As in Fig.~\ref{Fig.LocatErr_x_z1} and Fig.~\ref{Fig.LocatErr_x_z2}, the effective scope $w'$ in \eqref{equ:biscope} is depicted as each curve's scale. Whereas, the effective scope $l'$ is twice as each curve's scale in Fig.~\ref{Fig.LocatErr_y_z1} and Fig.~\ref{Fig.LocatErr_y_z2}. Therefore, we can conclude from the figures that the smaller the baseline is, the wider the effective scope is, meanwhile, the localization accuracy becomes more sensitive to the accuracy of beam spot's position estimation.

From the analysis above, the farther the target gets away from the BS and the less accuracy of spot position estimation, the more variation there will be. However, the localization errors in our simulation are generally a few centimeters. Hence, it can be concluded that the accuracy of BRBL is at the level of centimeters.


\subsection{Comparison}
\begin{figure}[!t]
	\centering
	\includegraphics[scale=0.55]{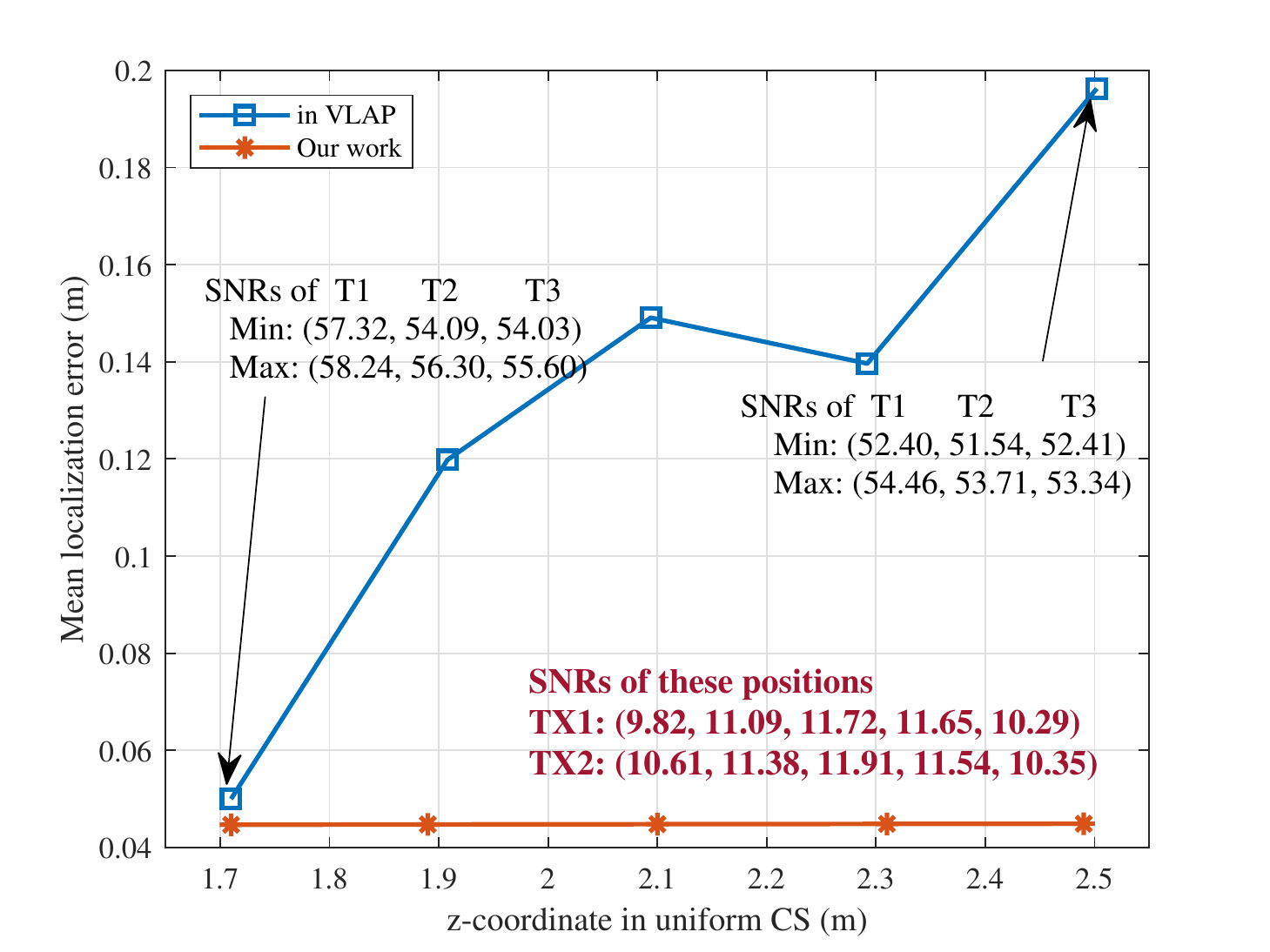}
	\caption{Comparisons between the visible light and accelerator based positioning system (VLCP) in~\cite{ComparisonAOA} and our work.}
	\label{Fig.Compare}
\end{figure}

In~\cite{ComparisonAOA}, an indoor positioning system using visible light and accelerator (VLAP) is proposed, of which the localization scheme is triangulation. The VLAP adopts three LEDs as transmitters, i.e., $T_1$, $T_2$ and $T_3$, while BRBL adopts two integrated transmitters, i.e., TX1, TX2. Considering that the WCS differs in VLAP and BRBL, we re-build a unified coordinate system (UCS). We put TX1 of BRBL and $T_1$ of VLAP together as the origin, and take the line connecting TX1 and TX2 as the $x$-axis, which is also the line connecting $T_1$ and $T_2$. Besides, the positive $z$-axis direction is set from ceiling to floor.

We perform the localization simulations as target's position changing in the UCS and present the errors in Fig.~\ref{Fig.Compare}. In these simulations, we fix the $x,y$-coordinates of target as $(0\mathrm{m},0.2\mathrm{m})$ and only vary $z$-coordinate from $1.7\mathrm{m}$ to $2.5\mathrm{m}$.
By comparing the $3$-D localization RMSE, the BRBL outperforms VLAP as the errors are around $4\mathrm{cm}$ and less fluctuation with different $z$-coordinates.

Furthermore, we also calculate the SNRs of multiple estimated positions to evaluate the required link conditions. According to the results provided in~\cite{ComparisonAOA}, the SNRs of each transmitter are almost $50-60\mathrm{dB}$, and gets slightly smaller as the target moves away from the ceiling. In BRBL, the SNRs are marked as bold fonts in Fig.~\ref{Fig.Compare}. The mean SNRs keep stable about $10\mathrm{dB}$, once the target is within the effective scope. Therefore, centimeter accuracy can be achieved without a very high SNR in BRBL.

\begin{table*}[h]
	\centering
	\caption{Comparison with existing visible light based ILSs.  }
	\label{tab:comparison}
		\resizebox{0.8\linewidth}{!}{
		\begin{threeparttable}
				\begin{tabular}{|c|c|c|c|c|c|c|}
					\hline
					\multicolumn{1}{|c|}{\bf{Technology}}
					&\multicolumn{1}{c|}{\bf{Ref.}}
					&\multicolumn{1}{c|}{\bf{Ceiling height}}
					&\multicolumn{1}{c|}{\bf{Accuracy}}
					&\multicolumn{1}{c|}{\bf{Note}}
					&\multicolumn{1}{c|}{\bf{Scheme}}\\ 
					\hline
					
					\multirow{5}*{Visible light} & \cite{li2018vlc}&  $0.8$ m & mean $6.58$ cm & 3 TXs & \multirow{3}*{\makecell[c]{Single-view\\vision analysis}}\\
					
					\cline{2-5} 
					\multirow{5}*{} & \cite{6817855}   & $3$ m &  $<10$ cm& 4 TXs&   \multirow{3}*{}\\
					
					\cline{2-5}
					\multirow{5}*{} &\cite{7855673} & $1.2$ m / $1.8$ m&\makecell[c]{$5.0$ cm @$1.2$ m \\$6.6$ cm @$1.8$ m }&3 TXs&\multirow{3}*{} \\
					
					\cline{2-5} \cline{6-6} 
					\multirow{5}*{} & \cite{rahman2011high}  &$3.5$ m& $\sim 10$ cm& 3 TXs &\multirow{3}*{\makecell[c]{Two-view\\vision analysis}} \\
					
					\cline{2-5} 
					\multirow{5}*{} & \cite{kim2012implementation}  & $2.7$ m &$8.5$ cm & 4 TXs &\multirow{3}*{}\\
					
					\cline{1-5} 
					Resonant beam & \makecell[c]{BRBL \\(this work)}&$1$ m / $2$ m & \makecell[c]{$<5$ cm @$1$ m \\$<13$ cm @$2$ m } &\makecell[c]{*2 TXs}&\multirow{3}*{}\\
					\hline
				\end{tabular}
			\begin{tablenotes}
				\footnotesize
				\item[*] For BRBL: sensors are in the TXs at BS.
			\end{tablenotes}		
		\end{threeparttable}
		}
	\end{table*}

Further, we provide a comparison table between BRBL and some ILSs to show the difference in characteristics and performances as Table~\ref{tab:comparison}. BRBL outperforms the others in some aspects, e.g., high accuracy with easy implementation, and battery-free target. As for precision, systems in \cite{li2018vlc,6817855,7855673} can achieve $<10 \mathrm{cm}$  accuracy at indoor scale, and require $3$ LEDs, $4$ LEDs and $3$ LEDs respectively. Systems in \cite{rahman2011high,kim2012implementation} integrate two image sensors in RX and require $3$ or $4$ reference LEDs for localization estimate. The accuracy of BRBL is at the same order of magnitude as other systems, but fewer luminaries are required. Moreover, the systems in \cite{li2018vlc,6817855,7855673,rahman2011high,kim2012implementation} share a feature that the RX integrates image sensors and thus needs energy supply in target, unlike the target in BRBL that is battery-free.





\section{Open Issues and Discussion}
In this section, we list some open issues that worth further study. We also provide some discussions and ideas as inspirations.

\subsection{Non Line of Sight (NLOS)}
Since BRBL is a one-to-one reflex structure system, the occlusion of line of sight (LOS) will cut off the only optical path, keep the two parties unconnected and make BRBL out of service. Nevertheless, the connection under NLOS state in RB-based systems can be reached, once the oscillation beam is generated in the NLOS paths. So the key point is to utilize relay nodes to ensure that there is at least one access point (AP) for the target when LOS is not satisfied. Solutions, such as deploying multiple BSs and placing multiple auxiliary reflectors~\cite{NLOS}, are applicable to expand paths of connection.

Considering that the position of target is unknown, applying fixed auxiliary reflector (relay node) may lead to unsuccessful localization. Hence, to increase the probability of successful relaying, like employing diverse reflective arrays or innovative reflective surfaces, is an effective approach. These proposals may mitigate or solve the NLOS issues for localization and are interesting topics for further study.

\subsection{Identification of Targets}
The premise of realizing identification of multiple targets in BRBL is that resonance can be generated between the TX in BS and RX in each target. In principle, it is available that one TX simultaneously links with multiple RXs in RBS. Hence, multiple targets can be positioned at the same time in BRBL as long as these targets are all within the effective scope. So the key of the target identification can be addressed as the distinction between the retroreflectors. In BRBL, the spot images are utilized for target localization. In this context, one possible direction is to number the TXs by means of coating the mirrors in the retroreflectors with different patterns, e.g., QR code. Thus, combining with the specific access protocol, the identification of targets as well as the access control can be obtained. Mechanism with multiple targets, coated reflective surfaces in RBS and access control are worth being investigated.


\subsection{Field of View}
As in Fig.~\ref{Fig.SNRdB}, when the target gets close to boundary of FoV of each TX, the SNR shows a significant decline. The FoV of each TX is approximate $22^\circ$ with $d=1\mathrm{m}$ and  $16^\circ$ with $d=2\mathrm{m}$. According to \eqref{equ:w} and \eqref{equ:biscope}, the effective scope of the BRBL is about $0.4\mathrm{m}\times0.4\mathrm{m}$ with $d=1\mathrm{m}$ and $0.6\mathrm{m}\times0.6\mathrm{m}$ with $d=2\mathrm{m}$, when $b$ approximately equals to $19 \mathrm{cm}$.

It should be noted that the FoV of single TX as well as the effective of the BRBL in this paper is a reference value under certain simulation parameters. We choose small system parameters for fast computation of the beam propagation simulation, for instance, cat’s eye radius and gain medium radius are $2.5\mathrm{mm}$. Deciding factors of the single TX's FoV include the gain from input power and the over-the-air transmission loss of resonant beam. In our simulation, the choices of system parameters such as cat’s eye radius and gain medium radius, are small to make computation fast and inexpensive.
Therefore, given a larger size of the cat’s eye, the FoV of single TX and effective scope of the BRBL will be improved to satisfy more application scenarios.

\subsection{Reducing Noise Effects}
According to the above analysis, to reduce the influence the noises and to improve the SNR in \eqref{equ:snrCal}, noise reduction can be employed during the photovoltaic conversion or the image processing afterwards. One way to reduce the noises resources, like dark current, is to use pinned photodiode (PPD) CMOS, and to adopt the correlated double sampling (CDS) circuits in the PPD CMOS. Besides, owing to the elimination of kTC noise and the inhibitory effect on low frequency noises, the SNR can be significantly improved in the system using PPD CMOS~\cite{8585674}. Moreover, employing image denoising algorithms, such as spatial domain filtering, transform domain filtering, variational denoising methods, CNN-based
denoising methods, etc., is also effective~\cite{7124798}. And the choice and corresponding effects are open issues worth further research.

\subsection{Statistical Analysis}
Since RBS is an emerging and promising technology, there is no well-accepted propagation model to describe the spatiotemporal characteristics of the RBS channel. In current RBS literature, the existing theoretical models include ``Field propagation model'' and ``Cyclic power model''. The former model is a self-consistent equation describing the transfer process of resonant beam but is a transcendental equation with no analytical solution. And the latter is a power model to describe the beam energy flow in resonator, the output of which is a quantity averaged over time and space. Neither can support a rigorous RBS channel description. Therefore, statistical analysis of the spatial and temporal effects of noises, such as the Cramer-Rao Bound (CRB) derivation, can be performed once a consensus model is reached.

\section{Conclusions}
\label{sec:con}
In this paper, we propose a high-accuracy localization scheme, BRBL, adopting the binocular method and resonant beam, which utilizes the energy-concentration and self-alignment of RBS to realize self-positioning targets with high-efficiency transmission link. Firstly, we build the binocular localization model, including the resonant beam transmission analysis and the geometric calculation of the binocular method, to exhibit the localization mechanism. Moreover, in order to analyze the accuracy of BRBL, we establish the cyclic power model of the resonant beam transfer, as well as the signal and noise models of beam spot imaging. Finally, BRBL exhibits a superior performance that can offer centimeter-level accuracy localization according to the numerical results.




%

\setcounter{figure}{0}





%



\ifCLASSOPTIONcaptionsoff
  \newpage
\fi



 
\bibliographystyle{IEEETran}
\small 
%
\bibliography{mybib}
%
%

%

%
%







\end{document}